\documentclass{statsoc}
\usepackage[a4paper]{geometry}

\usepackage{amsmath,amssymb}
\usepackage{mathtools}
\usepackage{algorithm}
\usepackage[noend]{algpseudocode}
\usepackage{subcaption}
\usepackage{graphicx}

\usepackage{hyperref}
\hypersetup{
	colorlinks=true,
	linkcolor=blue,
	filecolor=magenta,      
	urlcolor=cyan,
	citecolor = black 
}

\usepackage{natbib}

\title[Dynamic latent space REM]{Dynamic latent space relational event model}
\author{I. Artico}
\address{Universit\`a della Svizzera italiana, Lugano, Switzerland.}
\email{igor.artico@usi.ch}
\author[I. Artico and E.C. Wit]{E.C. Wit}
\address{Universit\`a della Svizzera italiana, Lugano, Switzerland.}
\email{ernst.jan.camiel.wit@usi.ch}

\begin{document}

\begin{abstract}
Dynamic relational processes, such as e-mail exchanges, bank loans and scientific citations, are important examples of dynamic networks, in which the relational events consistute time-stamped edges. There are contexts where the network might be considered a reflection of underlying dynamics in some latent space, whereby nodes are associated with dynamic locations and their relative distances drive their interaction tendencies. As time passes nodes can change their locations assuming new configurations, with different interaction patterns. 
		
The aim of this paper is to define a dynamic latent space relational event model. We then develop a computationally efficient method for inferring the locations of the nodes. We make use of the Expectation Maximization algorithm which embeds an extension of the universal Kalman filter. Kalman filters are known for being effective tools in the context of tracking objects in the space, with successful applications in fields such as geolocalization. We extend its application to dynamic networks by filtering the signal from a sequence of adjacency matrices and recovering the hidden movements. Besides the latent space our formulation includes also more traditional fixed and random effects, achieving a general model that can suit a large variety of applications.
\end{abstract}

%

\section{Introduction} \label{introduction}
Networks appear in many contexts. Examples include gene regulatory networks \citep{signorelli2016neat}, financial networks \citep{cook2014global}, psychopathological symptom networks \citep{de2017investigation}, political collaboration networks \citep{signorelli2018penalized}, and contagion networks \citep{uvzupyte2020test}. Studying networks is important for  understanding complex relationships and interactions between the components of the system. The analysis can be difficult due to the many endogenous and exogenous factors that may play a role in the constitution of a network. The aim of statistical modelling in this context is to describe the underlying generative process in order to assist in identifying drivers of these complex interactions. These models can assist in learning certain features of the process, filtering noise from the data, thereby making interpretation possible.



In this manuscript we are considering temporal random networks, whereby  nodes make instantaneous time-stamped directed or undirected connections. Examples are email exchanges, bank loans, phone calls, article citations. A common approach to these networks has been flattening the time variable and studying the resulting static network. Although this method simplifies the complexity of the calculations, clearly there is a loss of information about the temporal structure of the process. Most networks are  inherently dynamic. Subjects repeatedly create ties through time. Since the adjustment of ties is influenced by the existence and non-existence of other ties, the network is both the dependent and the explanatory variable in this process \citep{brandes2009networks}. 
Thus rather than viewing this as a static network, we consider the generative process as a  network structure in which the actors interact with each other through the time. Edges are defined as  instantaneous events. This quantitative framework is known as \emph{relational event modelling}.

The basic form of a relational event model as an event history model can be found in  \cite{butts20084} with an application to the communications during the World Trade Center disaster. The model has been extended by \cite{brandes2009networks} to weighted networks: nodes involved in these events are actors, such as countries, international organizations or ethnic groups. An event is assigned a positive or negative weight depending on a cooperative or hostile type of interaction, respectively. 
Other examples of relational event modelling include the work by  \cite{vu2017relational} on interhospital patient transfers within a regional community of health care organizations or the analysis of  social interaction between animals \citep{tranmer2015using}.

In a relational event model the connectivity may depend on the past evolution of the network. Keeping track of the past is challenging for dynamic networks because of the high number of possible configurations (k-stars, k-triangles, etc.) that could be taken into account, as well as their closure time and the time they keep affecting future configurations. 
We thus propose to take some kind of summary of the past configurations. A solution that can both summarize the process and approximate effectively the past information is the idea of a dynamic latent space.  
To describe the latent structure of a network one can think of placing the vertices in a space where the distance between two points describes the tendency or lack of tendency to connect. Among social scientists this is typically called a \emph{social space} where actors with more interactions are close together and vice versa \citep{bourdieu1989social}.
The locations are allowed to change in time. At each time point new connections are formed and  the subjects develop attraction/repulsion that force them to change their social space configuration.
The new configuration is the one that best reflect the new connectivity behavior.
As a result one location at a certain time reflects past information, within the limits of the latent space formulation.
This evolution describes the social history of the subjects, their preferences, and the groups they might join or leave.

The problem of tracking latent locations has been studied by many authors specifically for the static case, i.e., tracking locations under the assumption that they are fixed over time. For static binary networks \cite{hoff2002latent} provide a framework for inference. Some extensions of that model has been developed to overcome the limitations of the latent space formulation  \citep{hoff2005bilinear, hoff2008modeling, hoff2009multiplicative}. 

Similar to the latent space is the stochastic block model that  describes the similarity between the actors by grouping them together. An extension  of stochastic block modelling to relational event data is provided by  \cite{dubois2013stochastic}. 
An approach for modelling a latent space on dynamic binary networks was proposed by \cite{sarkar2005dynamic}. The method is based on a first preprocessing phase where raw location estimation are provided trough Multidimensional Scaling. In the estimation phase they treat the dynamic locations as fixed parameters and optimize them via a conjugate gradient approach.   The distances between nodes are approximated by cutting off the larger ones and including an additional penalty for forcing distant nodes to be closer. In our work, we aim to avoid making ad hoc assumptions.

\cite{sewell2015latent} developed a dynamic latent space with node specific parameters that regulate the incoming and outgoing links. The inference is performed via Metropolis Hastings algorithm. Instead, we use a Kalman filter, which is computationally more efficient. 
 
\cite{durante2016locally} developed a Bayesian model using a Polya-Gamma data augmentation for binary connections and Gaussian processes for parameter dynamics, with a non-Euclidean dissimilarity measure. Instead, we tackle the problem from a frequentist perspective providing a method which does not require data augmentation. Moreover, rather than embedding the dynamic latent space into a GLM, we embed it in a relational event model. Although non-Euclidean alternatives are possible, in our application we focus on an easily interpretable Euclidean latent space. Furthermore, our method can be applied to networks with non-binary links that are distributed according to any exponential family distribution.

In section~\ref{sec:models} we present several formulations of the latent space relational event model. In section~\ref{sec:inference} we propose an efficient inference method that is based on combing the state-space formulation of the model with the EM algorithm. In section~\ref{sec:applied} we check the performance and limitations of our method via simulations. In section~\ref{sec:patent} we aim to discover the latent structure of technological innovation, by studying over 23 million patent citations from 1967 until 2006.

\section{Latent space relational event models}
\label{sec:models}

In this section we introduce a general version of a latent space relational event model. We consider a set of actors, defined as a finite vertex set $V=\{1,\ldots,p\}$, that can exchange links or edges in time. In principle, we will consider the exchange of relational events, such as discrete interaction, e.g., sending an email or citing a patent, but we will also consider extensions to the quantitative exchanges, such as import and export. As drivers of the exchange process we consider both endogenous, such as reciprocity, and exogenous variables, such as vertex characteristics. One particular exogenous variable is the relative location of the vertices in some Euclidean latent space, which itself is defined as a dynamic process. 

We consider a non-homogeneous multivariate Poisson counting process $\mathbf{N}= \{N_{ij}(t)~|~i, j \in V, t\in [0,T]\}$ and a state-space process $\mathbf{X}=\{X_i(t)\in\mathbb{R}^d~|~t \in [0,T], i=1,\ldots, p\}$ relative to some standard filtration ${\mathcal F}$. In particular, we consider $\mathcal{F}$-measurable rate functions $\lambda_{ij}(t)$ that drive the components of the counting process. In particular, we assume that the rates $\lambda_{ij}(t)$ are functions of the underlying positions $X_i(t)$ and $X_j(t)$, besides possible other exogenous characteristics $B_{ij}(t)$ and endogenous features $\mathbf{N}(t)$,
\[ \lambda_{ij}(t) = g(d(X_i(t), X_j(t)),B_{ij}(t),N(t)),  \]
for some measurable function $g$. Two common choices for the way that the rate depends on the locations is either as function of the squared distance,
$$d(X_i(t),X_j(t))=||X_i(t)-X_j(t)||^2$$
or the relative activity dissimilarity 
$$d(X_i(t),X_j(t)) = \frac{<X_i(t),X_j(t)>}{||X_i(t)||}$$ 
between $i$ and $j$ \citep{hoff2002latent}. The former induces a symmetric interpretation, where the latter allows for a more complex asymmetric interpretation of the state-space. The interaction dynamics itself can be highly structured and parametrized, i.e., $g = g_\theta$, whereas the state-space dynamics is assumed to be a random walk at equally spaced time points $t^x_k$ in $[0,T]$,
\begin{equation} X_{t^x_k} = X_{t^x_{k-1}} + v_k, \label{eq:randomwalk} \end{equation}
with $v_k \sim N(0,\Sigma)$ and $t^x_0=0$. The covariance matrix
$\Sigma$ regulates the evolution of the latent process: a large variance allows longer jumps. Given the joint formulation  $(\mathbf{X},\mathbf{N})$ of the state-space and interaction process, we will assume that only the interaction process $\mathbf{N}$ is observed and the main aim of this paper is to infer the structure of the state-space $\mathbf{X}$ and the rate functions $\lambda$, or more specifically, the parameter $\beta$ associated with functional form $\lambda=g_\beta$.

Next, we will consider two particular special cases of the latent space formulation of the interacting point process defined above. First we consider the general case, in which the relational events are observed in continuous time. This is the traditional setting for relational events. We will also define a relational event model where the interactions can only happen at specific times. For example, bibliometric citations or patent citations only happen at prespecified publication dates. Furthermore, this model allows a generalization to non-binary relational events, such as export between countries, that can be dealt with in the same inferential framework.  
 
\subsection{Continuous time relational event process $\mathbf{N}$}
\label{sec:continuousN}

We consider a sequence of $n$ relational events, $\{(i_1,j_1,t_1), \ldots, (i_n,j_n,t_n)~|~ t_i\in [0,T], ~i,j \in V\}$ observed according to the above defined relational counting process $\mathbf{N}$. In a latent space relational event model, the rate is defined as 
\begin{equation}
	\label{eqn:lambda}
	\log\lambda_{ij}(t) = - d(X_i(t),X_j(t)) + f^G_{ij}(B_{ij}(t)) + f^D_{ij}( \{\mathbf{N}(\tau)|\tau < t\}). 
\end{equation}
where the latent space effect $d(X_i(t),X_j(t))$ that captures the ``vicinity'' of the actors. The drivers of the network dynamics can be of various type: \emph{exogenous effects},
$$f^G_{ij}(B_{ij}(t)) = \beta_G^t B_{ij}(t),$$ 
such as global covariates, node covariates, edge covariates, as well as \emph{endogenous effects}, 
$$f^D_{ij}( \{\mathbf{N}(\tau)|\tau < t\}) = \beta_D^t s(\{\mathbf{N}(\tau)|\tau < t\}),$$ 
where network statistics $s()$ capture endogenous quantities such as popularity, reciprocity, and triadic closure. The parameter vector $\beta$ determines the relative importance of the various effects. 

Conditional on the process $\mathbf{X}$, the distribution of the $l$th interarrival time $\Delta t_{ij,l} = t_{k_{ij,l}}-t_{k_{ij,l-1}}$ for  interaction $i\rightarrow j$  are generalized exponentials, with rates 
\[  \mu_{ij}(\Delta t_{ij,l}) = \int_{t_{k_{ij,l-1}}}^{t_{k_{ij,l}}} \lambda_{ij}(\tau)~d\tau, \] 
where $k_{ij,l}\in \{1,\ldots, n\}$ is the time indicator  of the $l$th occasion where $i\rightarrow j$ happened.

The full log-likelihood of the complete process $\{\mathbf{X}, \mathbf{N}\}$, can be factorized in two components,
\begin{equation}
	\label{eqn:lik-contN}
	\begin{split}
		l(\beta, \Sigma) =& \log p_\beta(\mathbf{N}|\mathbf{X}) + \log p_\Sigma(\mathbf{X}),
	\end{split}
\end{equation}
where $\log p_\Sigma(\mathbf{X})=  - \frac{n}{2} \log |\Sigma| 
-\frac{1}{2}\sum_{k}  {(x_k - x_{k-1})' \Sigma^{-1} (x_k - x_{k-1}) } $ and	$\log  p_\beta(\mathbf{N}|\mathbf{X}) =  -\sum_{k_{ij,l}} {\mu_{ij}(\Delta t_{ij,l})} + \log \lambda_{ij}(t_{k_{ij,l}})$, with observations $X(t_k^x) = x_k$.
Although it is common in the REM literature to simplify inference by using the partial likelihood, we keep the generalized exponential component, as it can be estimated more easily in the M-step of the EM algorithm, described in section~\ref{sec:inference}.

\subsection{Discrete time relational event process $\mathbf{N}$}
\label{sec:discreteN}

If the relational events are ``published'' only on prespecified discrete event times $\mathcal{T}= \{t^e_1, \ldots, t^e_n\}$, we will make an additional assumption that the rate $\lambda$ is constant with respect to the endogenous and exogenous variables inside the collection intervals $(t^e_k, t^e_{k+1}]$. In fact, with respect to the endogenous variable $\mathbf{N}$ it makes sense that no further information between the publication dates affects the rates. In other words,  assuming a log link for the hazard, for $t\in (t^e_k,t^e_{k+1}]$
\begin{equation}
	\label{eqn:lambda}
	\log\lambda_{ij}(t) = - d(X_i(t),X_j(t)) + f^G_{ij}(B_{ij}(t_k^e)) + f^D_{ij}( \{\mathbf{N}(\tau)|\tau \leq t^e_k\}). 
\end{equation}
As the interactions $i\rightarrow j$ are collected at $t_{k+1}^e$ from the observation intervals $(t^e_k, t^e_{k+1}]$, the resulting interval counts $$Y_{ij}(k)=N_{ij}(t^e_{k+1})-N_{ij}(t^e_k)$$ of the number of interactions between $i$ and $j$ are Poisson distributed with rate,
\[  \mu_{ij}(k) = \int_{t^e_k}^{t^e_{k+1}} \lambda_{ij}(\tau)~d\tau. \]


As long as the collection time process $\{t^e_k\}$ is finer than or equal to the change process $\{t^x_k\}$ of the latent process, we obtain a discrete-time relational event process, i.e., $\mu_{ij}(k) = (t_{k+1}^e-t_k^e)\lambda(t_k^e)$. An advantage of using discrete time is the reduction of the model complexity. It is not uncommon to observe thousands, even million of links. Such numbers are not surprising when we consider  $p(p-1)$ processes having an expected number of links $\mathbb{E}[\sum_{p(p-1)} N_{ij}(t)]$ that grows rapidly. For simplicity of notation we will assume that the relational event collection process and the jumps of the latent space are equal and unitary,
$$ \{ t_0^x = t_0^e = 0,~~t_1^x = t_1^e = 1,\ldots, t_n^x = t_n^e = T  \}.$$
The model can be written as a discrete-time state space process,
\begin{equation}
\label{eqn:ss}
\begin{cases}
    X_k = X_{k-1} +  v_k \\
    Y_{ij}(k) \sim \mbox{Poi}(\mu_{ij}(k)),~~ 1\leq i \neq j\leq p \ 
\end{cases}
\end{equation}
where $v_k \sim N(0, \Sigma)$. 
\begin{figure}
	\centering
	\includegraphics[width=0.5\textwidth]{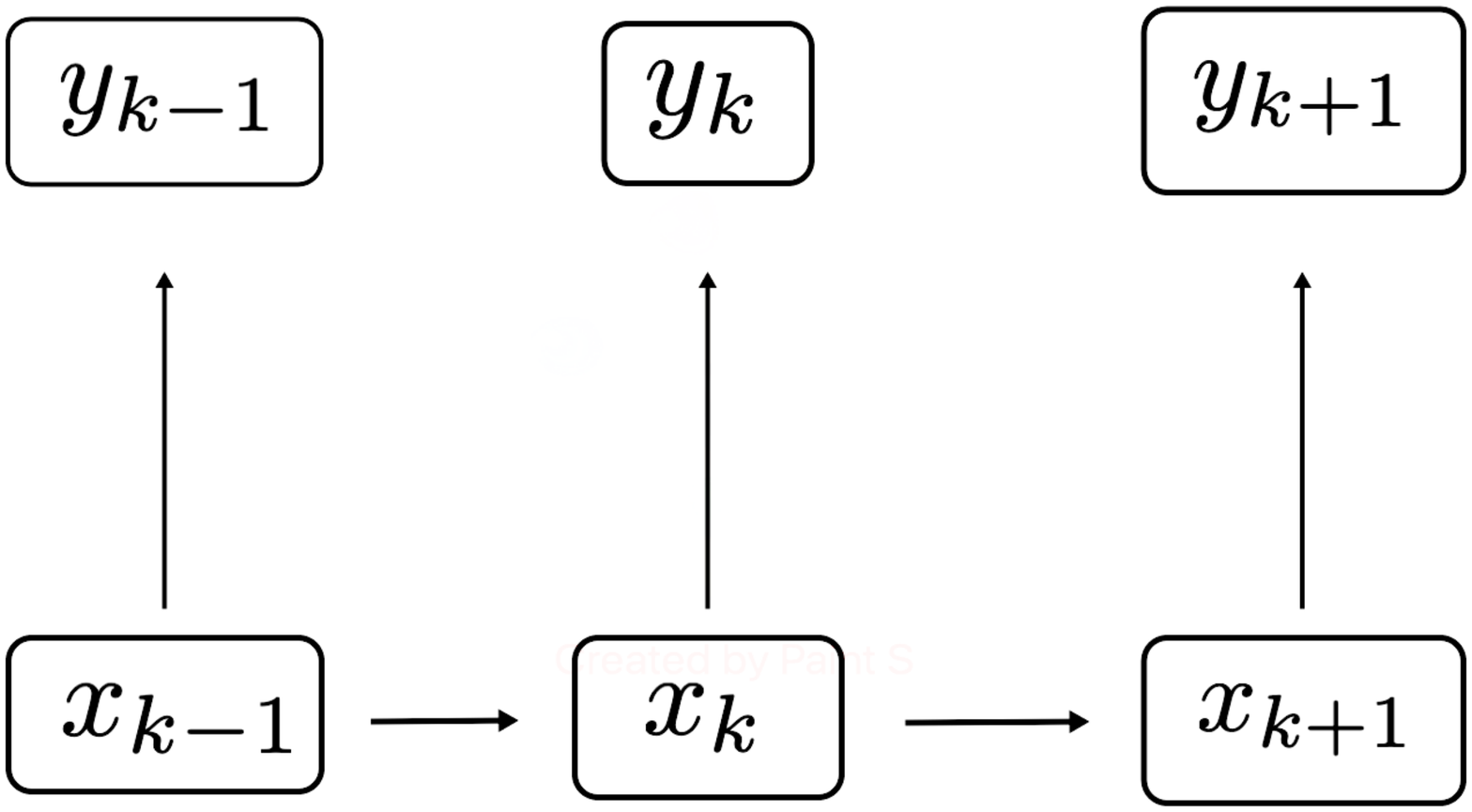} 
	\caption{The observed counts $y_k$ are a result of the dynamics in nodes locations $x_k$. Hence, $y$ is independent conditionally to the latent locations $x$.}
	\label{fig:process}
\end{figure}
 Given the observations $Y=y$ and $X=x$, the complete log-likelihood for the state space model in (\ref{eqn:ss}) can again be factorized in two components,
 \begin{equation}
 	\label{eqn:lik-discN}
 	\begin{split}
 		l(\beta, \Sigma) =& \log p_\beta(\mathbf{Y}|\mathbf{X}) + \log p_\Sigma(\mathbf{X}),
 	\end{split}
 \end{equation}
 where $\log p_\beta(\mathbf{Y}|\mathbf{X}) = -\sum_{kij} {\mu_{ij}(k)} + \sum_{kij} y_{ij}(k) \log \mu_{ij}(k) $  and $\log p_\Sigma(\mathbf{X})$ as above, where the factorization is according to the directed graph in Figure \ref{fig:process}, where $y_k \perp y_{-k},x_{-k}|x_k$ and $x_{k+1} \perp x_{k-1}|x_{k}$. Similar to \cite{butts20084} and \cite{perry2013point}, who focused on non-homogeneous exponential waiting times, this  approach focuses on non-homogeneous Poisson counts.

One advantage of the latent space formulation is the dimensionality reduction in the latent representation. As the number of nodes $p$ increases the number of observed counts $p(p-1)n$ grows quadratically while the latent space grows linearly as $pdn$.

\paragraph{Dynamic exponential family network model.}
\label{sec:extensions}
Given the state space formulation in (\ref{eqn:ss}), it is possible to generalize the model considering connections drawn from any exponential family distribution without changing the inference procedure. In fact, ignoring the connection with any underlying counting process, we could define a temporal network process on discrete time intervals $k$ ($k\in\{1,\ldots,n\})$ between nodes $i$ and $j$ as $ f(y_{ij}(k)) = \exp((y_{ij}(k)\theta -b(\theta))/a(\varphi) + c(y_{ij}(k),\varphi)$, where $\theta$ is the edge-specific canonical parameter. Using the canonical link function, we can specify the canonical parameter in a similar fashion to (\ref{eqn:lambda}),
\[ \theta(x_k) =  - d(x_i(k),x_j(k)) \]
where the values for $x$ are the latent states as before. It is also possible to add additional covariates, but we do not consider this case here. 
The inferential method presented in this manuscript remains mostly the same with a minimal change, effectively replacing the mean $\mu(x_k)$  and variance $R_k$ of the process by 
$$\mu(x_k) = b'(\theta)|_{x_k} \mbox{ and } R_k=b''(\theta)a(\varphi)|_{x_k}.$$
This generalized temporal network model can be used to model import and export or other dynamic networks with weighted edges. 


\section{Inference}
\label{sec:inference}

In this section we develop all the necessary steps for making inference on the latent states $x_k$ and the parameters $\Sigma$ an $\beta$.
Since the latent process $x_k$ is unobserved we aim to maximize $\int_x L(\beta, \Sigma ; y,x) dx$. We use the Expectation Maximization (EM) algorithm \citep{dempster1977maximum}. EM algorithm is widely used in problems where certain variables are missing or latent. The EM algorithm consists of an iterative maximization of the conditional expectation of the latent process $\mathbf{X}|\mathbf{N}, \beta,\Sigma$ with respect to the data.

Due to the stepwise dynamic of the latent locations (\ref{eq:randomwalk}) the expectation step is equivalent for  both  models presented in Section (\ref{sec:continuousN}) and Section (\ref{sec:discreteN}). As the locations are constant within intervals $\mathcal{T}$, the continuous time non-homogeneous exponential relational event model $\mathbf{N}$ reduces to a discrete time Poisson model $\mathbf{Y}$ during the E-Step.

\begin{equation}
	\notag
	Q(\beta,\Sigma|\beta^*,\Sigma^*)= \mathbb{E} [l_\mathbf{X}(\beta, \Sigma)|\mathbf{y}]. 
\end{equation}
where $\beta^*,\Sigma^*$ denote the parameters estimated at the previous EM iteration.
In the maximization step $Q(\beta, \Sigma|\beta^*,\Sigma^*)$ is maximized 
with respect to the parameters $\beta, \Sigma$.
The two steps above are iterated until convergence is reached.
The expectation step is typically challenging due to the high dimensional nature of the integral.

The expectation of the log-likelihood can approximately be written as a function of the first two conditioned moments $\mathbb{E}[x_k | y_{1:n}]$ and $\mathbb{V}[x_k | y_{1:n}]$. Exploiting the state space formulation of the model (\ref{eqn:ss}) we can estimate these two quantities with a Kalman filter and smoother \citep{kalman1960new}.  
The filter derives mean and variance of the latent process $x_k$ conditioned to the information on $y$ up to time k,
\begin{equation}
\notag
\begin{split}
\hat{x}_{k|k} =  \mathbb{E}[x_k | y_{1:k}] \qquad
V_{k|k} =        \mathbb{V}[x_k | y_{1:k}].
\end{split}
\end{equation}
The smoother refines these quantities accounting for the complete information on $y$ up to time $n$,
\begin{equation}
\notag
\begin{split}
\hat{x}_{k|n} =  \mathbb{E}[x_k | y_{1:n}] \qquad
V_{k|n} =        \mathbb{V}[x_k | y_{1:n}].
\end{split}
\end{equation}
The expected log-likelihood can be then calculated using these quantities obtained from the smoother. 
\begin{figure}[t]
	\centering
	\includegraphics[width=0.7\textwidth]{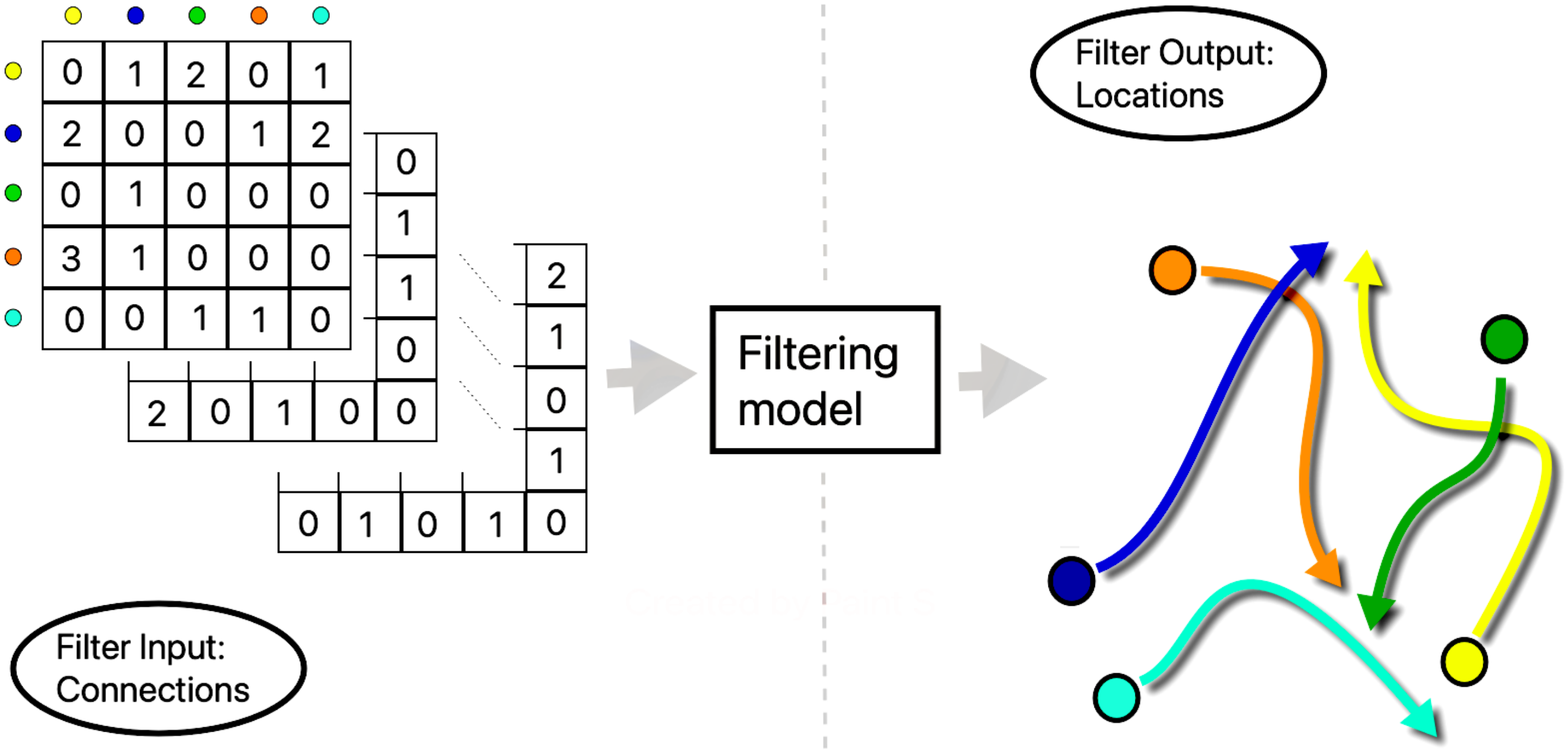} 
	\caption{The filtering model takes as input a sequence of adjacency matrices and update the node locations in the latent space.}
	\label{fig:filter}
\end{figure}

\subsection{E-Step: Extended Kalman Filter}
The Kalman filter is one of the most popular algorithms for making inference on state space models and it provides a solution that is both computationally cheap and accurate. Kalman filter is an iterative method that calculates the conditional distribution of the latent $x_k$. Given the causal DAG at Figure (\ref{fig:process}) $x_k$ depends on $x_{k-1}$ and the observed $y_k$.  Assuming a prior knowledge on the distribution of $x_{k-1}$ the conditional distribution of $x_k$ is calculated easily. The procedure is applied sequentially from time 1 to $n$, where the conditional distribution achieved at time $k$ becomes the prior knowledge for the next time point. An arbitrary distribution is specified for the initial $x_0$.
Calculating the conditional distribution entirely could be difficult so the first moments are calculated only.
The calculation of the conditional probability involves two steps that are universal in the filtering literature: predict and update. In order to be consistent to the forementioned literature we denote $\hat{x}_{k|k}=\mathbb{E}[x_{k}|y_{1:k}]$ and $V_{k|k}=\mathbb{V}[x_{k}|y_{1:k}]$ as the expectation and variance conditioned of having observed $y_k$.

\subsubsection*{Predict}
Assume that at time $k-1$ the approximated conditional distribution of the latent locations is $x_{k-1|k-1} \sim N(\hat{x}_{k-1|k-1}, V_{k-1|k-1})$.
For the initial case $k=1$ we set arbitrarily $x_{0|0}=v_0$ and $V_{0|0}=\Sigma_0$.
The predict step calculates the first moments of $x_k$ conditioned to $y_{k-1}$. In fields such physics, chemistry or engineering it is common to employ a forward function $x_k = f(x_{k-1}) + v_k$ which is related to the physical properties of the system.
In our case the random walk formulation makes no constraints on the latent process evolution. The forward function is the identity with moments
\begin{equation}
\notag
\begin{split}
&    \hat{x}_{k|k-1} = \mathbb{E}[x_{k-1} + v_k|y_{1:k-1}] = \hat{x}_{k-1|k-1} \\
&    V_{k|k-1} = \mathbb{V}[x_{k-1} + v_k|y_{1:k-1}] = V_{k-1|k-1} + \Sigma 
\end{split}        
\end{equation}
These are called the apriori mean and variance of the latent locations before observing $y_k$. The prior distribution is $x_{k|k-1} \sim N(\hat{x}_{k|k-1}, V_{k|k-1})$.
\begin{figure}[t]
	\centering
	\includegraphics[width=0.5\textwidth]{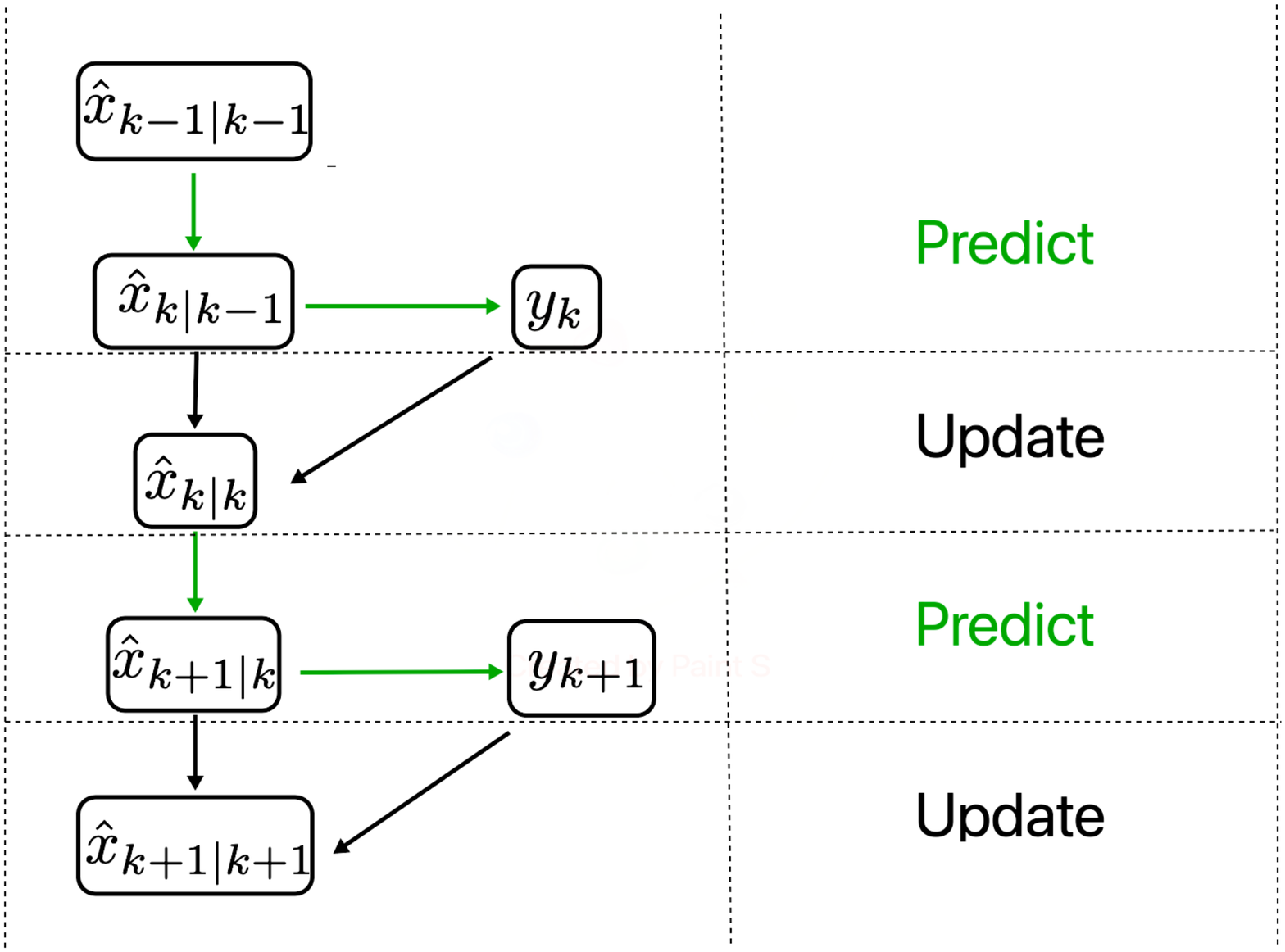} 
	\caption{The filtering procedure can be summarized as a sequence of predictions and updates. At each time step a prediction on the observed links count is made. The prediction error is then propagated back to the nodes for updating their positions.}
	\label{fig:ss}
\end{figure}

\subsubsection*{Update}
\label{sec:update}
The update step finalizes the calculation of the conditional distribution.
We consider $\mathbb{V}[y_k] = R_k $ where counts are independent with variance equal to the mean $R_k = \mu(x_k,  \beta) \hspace{0.005 \textwidth} \mathbb{I}_{p_y}$. In case a general dynamic network model using exponential family weighted edges, as described in Section (\ref{sec:extensions}), is considered then the mean $\mu(x_k)$ and variance $R_k$ vary accordingly.

Kalman filters assume that the observed process $y_k$ is Gaussian and the transformations involved are linear. The Extended Kalman Filter \citep{anderson2012optimal} overcomes the Kalman filter limitations.
By means of a first order Taylor expansion
\begin{equation}
\label{eqn:taylor}
\mu(x_k,  \beta) = \mu(\hat{x}_{k|k-1}, \beta) + H_k (x_k-\hat{x}_{k|k-1}), \qquad  H_k = \frac{\partial \mu(x, \beta) }{\partial x} \big|_{\hat{x}_{k|k-1}}
\end{equation}
we calculate the expectation $\mathbb{E}[y_k | y_{k-1}] =  \mu(\hat{x}_{k|k-1}, \beta)$, variance $\mathbb{V}[y_k | y_{k-1}] = H_k V_{k|k-1} H_k' + R_k$ and covariance $\mathbb{C}ov[x_k, y_k | y_{k-1}] =   V_{k|k-1} H_k'$ of the conditional predictive distribution of $y_k$.

The joint multivariate distribution of the observed and latent process is
\begin{equation}
\notag
\begin{bmatrix}
x_{k} \\
y_k
\end{bmatrix}
\big|y_{1:k-1} 
\sim \mathcal{L}\left(
\begin{bmatrix}
\hat{x}_{k|k-1} \\
\mu(\hat{x}_{k|k-1}, \beta)
\end{bmatrix}
,
\begin{bmatrix}
V_{k|k-1} & H_k V_{k|k-1} \\
V_{k|k-1} H_k' & H_k V_{k|k-1} H_k' + R_k
\end{bmatrix}
\right)  
\end{equation}
where $\mathcal{L}$ is some probability law parametrized by the first two moments.
Using the multivariate regression formulation we have the conditional moments of $x_k$
\begin{equation}
\label{filteringMatrix}
\begin{split}
    \hat{x}_{k|k} &= \mathbb{E}[x_k|y_{1:k}] = \hat{x}_{k|k-1} + K_k (y_k - \mu(\hat{x}_{k|k-1}, \beta) ) \\
    V_{k|k} &= \mathbb{E}[(x_k-\hat{x}_{k|k})(x_k-\hat{x}_{k|k})'|y_{1:k}] = (\mathbb{I} - K_k H_k ) V_{k|k-1}, \\
    K_k &= 
    V_{k|k-1} H_k' (R_k + H_k V_{k|k-1} H_k')^{-1},  
\end{split}
\end{equation}
see at Appendix \ref{A} for more details. We hence obtain posterior distribution $x_{k|k} \sim N(\hat{x}_{k|k}, V_{k|k})$, which is approximated to be Gaussian. This will be  the starting distribution for the inference at time $k+1$.
The filtering procedure is shown in Algorithm \ref{EKF}.
In Figure \ref{fig:filter} we show a visual representation of the algorithm: at each time point the model takes as input an adjacency matrix and returns the locations in the latent space.

In the update step the latent locations are updated according to the magnitude of the prediction error: a larger error in the prediction corresponds to a wider change in the locations. The filtering matrix $K_k$,  capturing the linear relationship between the latent and observed processes, weights this prediction error. $K_k$ is the ratio between the noise $R_k$ and the latent variance $\Sigma$. Thus $K_k$ filters the prediction error according to the signal/noise ratio. 
\cite{fahrmeir1992posterior} simply consider it as a single Fisher Scoring step, see Appendix \ref{D}.

\begin{algorithm}[t]
\caption{\textit{Extended Kalman Filter} }\label{EKF}
\begin{algorithmic}
\item \textit{Initialize} $\hat{x}_{0|0} = v_0$ and $V_{0,0} = \Sigma_0$
\For{k = 1, \dots, n}
\begin{enumerate}
    \item \textit{Filter prediction step}
    \begin{itemize}
        \item[] $\hat{x}_{k|k-1} = \hat{x}_{k-1|k-1}$  
        \item[] $V_{k|k-1} = V_{k-1|k-1} + \Sigma$
    \end{itemize}  
    \item \textit{Filter update step}
    \begin{itemize}
        \item[] $\hat{x}_{k|k} = \hat{x}_{k|k-1} + K_k(y_k - \mu(\hat{x}_{k|k-1}, \beta) )$  
        \item[] $V_{k|k} = (I - K_k H_k)V_{k|k-1}$
    \end{itemize}
    \textit{where}
    \begin{itemize}
        \item[] $K_k = V_{k|k-1}H_k^{'} (H_k V_{k|k-1} H_k^{'} + R_k)^{-1}$
        \item[] $H_k =  \frac{\partial \mu(x, \beta) }{\partial x} \big|_{\hat{x}_{k|k-1}}$
        \item[] $R_k = \mu(\hat{x}_{k|k-1}, \beta) \hspace{0.005 \textwidth} \mathbb{I}_{p_y}$
    \end{itemize}
\end{enumerate}    
\EndFor
\end{algorithmic}
\end{algorithm}

The Kalman filter can be interpreted as both a frequentist and Bayesian method. Under a Bayesian perspective the filtering procedure consists of a sequence of updates of the posterior mean and variance \citep{gamerman1991dynamic, gamerman1992dynamic, west1985dynamic}.
From the frequentist side the estimation based on the posterior mode is equivalent to the maximization of a penalized likelihood  \citep{fahrmeir1991kalman, fahrmeir1992posterior}, see Appendix \ref{D}. Approximating the posterior distribution with the same family of the prior, i.e., Gaussian, the posterior mean is equivalent to the posterior mode and hence the equivalence of the two approaches. This double interpretation makes Kalman filters appealing for both types of applications.

\subsection*{Smoother}
The smoother moves backward from the last prediction to the first.  It calculates the first moments of the latent process conditioned to the information of all time points.

Similarly as the EKF, the backward matrix $B$ can be calculated considering the multivariate distribution of the latent locations at two consecutive time points, 
\begin{equation}
\notag
\begin{bmatrix}
x_{k-1} \\
x_k
\end{bmatrix}
\big|y_{1:k-1} 
\sim N\left(
\begin{bmatrix}
\hat{x}_{k-1|k-1} \\
\hat{x}_{k|k-1}
\end{bmatrix}
,
\begin{bmatrix}
V_{k-1|k-1} & V_{k-1|k-1} \\
V_{k-1|k-1} & V_{k|k-1}
\end{bmatrix}
\right)  .
\end{equation}
Using the multivariate regression formula we have the conditioned mean of $x_{k-1}$ over $x_{k}$
\begin{equation}
\notag
\mathbb{E}\left[ x_{k-1} | x_k, y_{1:k-1} \right] = \hat{x}_{k-1|k-1} + B_k (x_k - \hat{x}_{k|k-1}) \quad \text{with} \quad   B_k = 
V_{k-1|k-1} V_{k|k-1}^{-1}
\end{equation}
According to the conditional independence in Figure (\ref{fig:process}) we have $(x_{k-1} \perp y_{k:n}) |x_{k}$ since $x_k$ closes the dependency path. Using the iterated expectation rule we have
\begin{equation}
\notag
\begin{split}
 \hat{x}_{k-1|n} &= \mathbb{E}\left[x_{k-1} |y_{1:n} \right]  = \mathbb{E}\left[\mathbb{E}\left[ x_{k-1} | x_k, y_{1:n} \right] | y_{1:n} \right]  = \mathbb{E}\left[\mathbb{E}\left[ x_{k-1} | x_k, y_{1:k-1} \right] | y_{1:n} \right] \\
 &= \mathbb{E}\left[ \hat{x}_{k-1|k-1} + B_k (x_k - \hat{x}_{k|k-1}) | y_{1:n} \right]  \\
 &= \hat{x}_{k-1|k-1} + B_k (\hat{x}_{k|n} - \hat{x}_{k|k-1})
\end{split}
\end{equation}
where $\hat{x}_{k-1|k-1} $ and $ \hat{x}_{k|k-1} $  are constants. 
In the same way using the iterated variance rule
\begin{equation}
\notag
\begin{split}
\mathbb{V}\left[x_{k-1} |  y_{1:n} \right] &= \mathbb{E}\left[\mathbb{V}\left[x_{k-1} |x_{k}, y_{1:n} \right] | y_{1:n}\right] +
\mathbb{V}\left[\mathbb{E}\left[x_{k-1} |x_{k}, y_{1:n} \right] | y_{1:n}\right] \\
&= V_{k-1|k-1} - B_k V_{k|k-1} B_k' + B_k V_{k|n} B_k' \\
&= V_{k-1|k-1} + B_k(V_{k|n} - V_{k|k-1} )B_k',
\end{split}
\end{equation}
see at Appendix \ref{B} for more details. The smoothing procedure is presented in Algorithm \ref{smoother} and it is known as the Rauch-Tung-Striebel smoother. The final iteration of the smoother updates the starting values $\hat{x}_{0|0}$ and $V_{0|0}$. These values will be used as starting points for the successive EM iteration.
\begin{algorithm}[t]
\caption{\textit{Smoother} }\label{smoother}
\begin{algorithmic}
\For{k = n, \dots, 1}
\begin{enumerate}
    \item \textit{Backward step}
    \begin{itemize}
        \item[] $\hat{x}_{k-1|n} = \hat{x}_{k-1|k-1} + B_k(\hat{x}_{k|n} - \hat{x}_{k|k-1} )$
        \item[] $V_{k-1|n} = V_{k-1|k-1} + B_k(V_{k|n} - V_{k|k-1} )B_k^{'}$
    \end{itemize}
    \textit{where}
    \begin{itemize}
        \item[] $B_k = V_{k-1|k-1} V_{k|k-1}^{-1}$
    \end{itemize}
\end{enumerate}    
\EndFor
\end{algorithmic}
\end{algorithm}

\subsection{M-Step: a Generalized Additive Model}
\label{sec:GAM}
In the maximization step we maximize the log-likelihood with respect to the parameters $\beta, \Sigma$ and we make the first distinction between the continuous (\ref{eqn:lik-contN}) and discrete (\ref{eqn:lik-discN}) time models.
For the continuous time process $\mathbf{N}$ the expected log-likelihood is
\begin{equation}
		Q^{\mathbf{N}}(\beta,\Sigma|\beta^*,\Sigma^*) = \mathbb{E} [\log p_\beta(\mathbf{N}|\mathbf{X})|y_{1:n}] + \mathbb{E} [ \log p_\Sigma(\mathbf{X})|y_{1:n}] = Q^P(\beta) + Q^G(\Sigma).
\nonumber
\end{equation}
For the discrete time process $\mathbf{Y}$ the expected log-likelihood is
\begin{equation}
		Q^{\mathbf{Y}}(\beta,\Sigma|\beta^*,\Sigma^*) = \mathbb{E} [\log p_\beta(\mathbf{Y}|\mathbf{X})|y_{1:n}] + \mathbb{E} [ \log p_\Sigma(\mathbf{X})|y_{1:n}]= Q^E(\beta) + Q^G(\Sigma).
\nonumber
\end{equation}
Notice that the Poisson component $Q^P(\beta)$ and exponential component $Q^E(\beta)$ do not depend on $\Sigma$ as well as the Gaussian component $Q^G(\Sigma)$ does not depend on the remaining parameters $\beta$. These quantities can be optimized separately.

\subsubsection*{Gaussian component}
We can maximize the Gaussian component 
\begin{equation}
\notag
       Q^G(\Sigma)=
      -\frac{1}{2} \sum_{k=1}^n \mathbb{E} [(x_k - x_{k-1})' \Sigma^{-1} (x_k - x_{k-1})  |y_{1:n}]  -n\log|\Sigma| - \frac{n}{2} \log(2\pi).
\end{equation}
finding the zero of the first derivative with respect to $\Sigma$.
Rearranging the elements and taking the expectation as shown in Appendix \ref{C} we obtain
\begin{eqnarray*}
\hat{\Sigma} &=& \mathbb{E}\left[ \frac{1}{n} \sum_1^n  (x_k - x_{k-1})(x_k - x_{k-1})' \big|y_{1:n} \right] \\
&=& \frac{1}{n} \sum_1^n  V_{k|n} + V_{k-1|n} + B_k V_{k|n} +  V_{k|n}B_k' + (\hat{x}_{k|n} - \hat{x}_{k-1|n})(\hat{x}_{k|n} - \hat{x}_{k-1|n})'
\end{eqnarray*}
This result corresponds to the one presented in \cite{fahrmeir1994dynamic}.
Substituting  $V_{k|n}B_k' = \mathbb{C}\text{ov}(x_{k|n}, x_{k-1|n} \big|y_{1:n}) $ we have the equivalence with the result of \cite{watson1983alternative}.


It is crucial to have a good estimate $\Sigma$. Having $\Sigma$ small implies that a little portion of the prediction error is used to update the locations and therefore the latent process moves slowly and delayed. When $\Sigma$ is high the estimated latent process is heavily influenced by the last observation and have a tendency to overfit the observed process. In some practical fields $\Sigma$ is tuned manually by searching for overfitting or delayed behaviors in the errors. Our EM provides a precise solution and avoid the manual tuning.

\subsubsection*{Poisson component}
For arbitrary exponential family distributed edges, as described in Section (\ref{sec:extensions}), the observed process component can be maximized numerically with a general optimization algorithm. However, for Poisson distribution a more elegant solution is available. 
The expectation of the Poisson component for the discrete time process $\mathbf{Y} $ can be  rearranged as follows
\begin{equation}
\notag
    \begin{split}
       Q^P(\beta)&=\sum_{tij} \mathbb{E}[-\mu_{ij}(x_k, \beta) + y_{ij}(k)\log(\mu_{ij}(x_k, \beta)) - \log(y_{ij}(k)!)|y_{1:n}] \\
        &= \sum_{tij} - \mu_{ij}^*(x_k, \beta) + y_{ij}(k)\log(\mu_{ij}^*(x_k, \beta)) - \log(y_{ij}(k)!) + C
    \end{split}
\end{equation}
where, up to an additive constant, the expected log-likelihood can be formulated as a Poisson log-likelihood with the associated rates
\begin{equation}
\label{eq:offset}
\log(\lambda_{ij}^*(x_k, \beta)) = \log(\mathbb{E} [e^{- d(x_i(k),x_j(k))}|y_{1:n}]) + f^G_{ij}(B_{ij}(t_k^e)) + f^D_{ij}( \{\mathbf{N}(\tau)|\tau \leq t^e_k\}).
\end{equation}
The optimization can be performed by fitting a Generalized Additive Model (Wood, 2013) with this linear predictor and the offset $\log(\mathbb{E} [e^{- d(x_i(k),x_j(k))}|y_{1:n}])$. See  Appendix \ref{C} for the full derivation.

The  expected value in the offset cannot be further simplified. We use a second order Taylor approximation, which can be expressed as a function of the first two moments of the latent locations, $\mathbb{E}[x_k | y_{1:n}]$ and $\mathbb{V}[x_k | y_{1:n}]$. Consider $g_{i,j}(x)=e^{- d(x_i(k),x_j(k))}$, then the expectation of the Taylor expansion at $x_{k|k}$ is 
\begin{equation}
\label{eqn:TaylorEM}
\begin{split}
    \mathbb{E} [g_{i,j}(x)|y_{1:n}] 
    &= g_{i,j}(x_{k|k}) + \frac{1}{2}\text{trace}\left(\frac{\partial^2 g_{i,j}(x)}{\partial^2 x}\big|_{x_{k|k}} V_{k|k} \right), 
\end{split}
\end{equation}
where the expectation of the first derivative term is zero.

The GAM model is an elegant way to specify the remaining fixed and random effects. This formulation is very general and allows to estimate constant and linear effects or to use splines for estimating non-linear and time-varying effects.

\subsubsection*{Exponential component}

The expectation of the exponential component for the continuous time  process $\mathbf{N}$ is
\begin{equation}
\notag
       Q^E(\beta)=  \mathbb{E}\left[  -\sum_{k_{ij,l}} {\mu_{ij}(\Delta t_{ij,l})} + \log \lambda_{ij}(t_{k_{ij,l}}) |y_{1:n} \right]
\end{equation}
Note that, up to a multiplicative constant $y_{ij}(k)$, the exponential log-likelihood factorizes similarly to that of the Poisson. Even in this case the expected log-likelihood can be rewritten as an exponential log-likelihood with the same offset as (\ref{eq:offset}). The inference is performed via survival regression with rates
\begin{equation}
\notag
\log(\lambda_{ij}^*(x_k, \beta)) = \log(\mathbb{E} [e^{- d(x_i(k),x_j(k))}|y_{1:n}]) + f^G_{ij}(B_{ij}(t)) + f^D_{ij}( \{\mathbf{N}(\tau)|\tau < t\})
\end{equation}
and exponential waiting times.

\begin{algorithm}[t]
\caption{\textit{Expectation Maximization} }\label{EM}
\begin{algorithmic}
\item \textit{Initialize} $\hat{x}_{0|0} = v_0$, $V_{0|0} = \Sigma$, 
$\Sigma = \Sigma_0$ and $\beta = \beta_0$
\While{not converged}
\begin{enumerate}
    \item Expectation:
    \begin{itemize}
        \item[-]  \textit{Extended Kalman Filter}
        \item[-]  \textit{Smoother}
    \end{itemize}
    \item Maximization and update of starting values:
    \begin{itemize}
        \item[] $\beta = \operatorname*{argmax}_\beta Q(\beta)$ 
        \item[] $\Sigma = \hat{\Sigma}$
        \item[] $\hat{x}_{0|0} = \hat{x}_{0|n}$
        \item[] $ V_{0|0} = V_{0|n}$
    \end{itemize}
    \item Check for convergence  
\end{enumerate}    
\EndWhile
\end{algorithmic}
\end{algorithm}

\subsection{Higher order approximation}

The EKF is based on a first order Taylor expansion in (\ref{eqn:taylor}).
We can approximate the $\mu$ function with a order higher. A popular solution is the Unscented Transformation, the key solution of the Unscented Kalman Filter (UKF)  \citep{julier1996general, 10.1117/12.280797}. 
The algorithm has a similar shape as the EKF with the difference that the filtering matrix $K_k$ is calculated empirically. 
We begin with a fixed number of points to approximate a Gaussian by creating a discrete distribution having the same first and second (and possibly higher) moments. Each point in the discrete approximation can be directly transformed. The mean and the covariance of the transformed ensemble can then be computed as the estimate of the nonlinear transformation of the original distribution.

Given a $pd$-dimensional Gaussian having covariance $V_{k|k-1}$ we can construct a set of points having the same sample covariance from the columns (or rows) of the matrices  $\sqrt{(\kappa + pd)V_{k|k-1}}$. The square root of the matrix is typically done via a  Cholesky decomposition. Adding and subtracting these points to $\hat{x}_{k|k-1}$ yields a symmetric set of $2pd+1$ points (central point included) having the desired sample mean and covariance. This is the minimal number of points capable of encoding this information \citep{julier1996general}. We then calculate the sample mean and covariance of the transformed points. Finally, the filtering matrix $K_k$ can be calculated as the rate between the sample covariance and the sample variance. 
\begin{equation}
\notag
\begin{split}
&    K_k = \widehat{\mathbb{C}ov}(x_{k}, y_k |y_{1:k-1}) \widehat{\mathbb{V}}(y_k |y_{k-1})^{-1}.   
\end{split}
\end{equation}
The Unscented Kalman Filter is presented in Algorithm \ref{UKF}. The prediction and the update step are the same as those of the EKF.
The $\kappa$ parameter regulates both the weight of the central point and the spreading of the other points: a large $\kappa$ leads to a wider spreading of the points.  \cite{10.1117/12.280797} suggests a useful heuristic to select $pd + \kappa = 3$. The use of the Unscented Kalman filter makes the computation of (\ref{eqn:TaylorEM}) straightforward by simply taking the sample mean of the transformed ensemble.  

\begin{algorithm}[t]
\caption{\textit{Unscented Kalman Filter} }\label{UKF}
\begin{algorithmic}
\item \textit{Initialize} $\kappa = \kappa_0$ 
 \item[] $w_0 =  \kappa/(p_x +\kappa)$ 
  \item[] $w_j = 1/2(p_x +\kappa), \quad j=1, \dots, 2p_x$
\For{k = 1, \dots, n}
\begin{enumerate}
    \item \textit{Filter prediction step}
    \item \textit{Filtering matrix calculation}
    \begin{itemize}
        \item[] $A = V_{k, k-1}^{\frac{1}{2}}$
        \item[] $s_0 = \hat{x}_{k, k-1}$
        \item[] $s_j = \hat{x}_{k, k-1} + \sqrt{pd +\kappa} A_j, \quad j=1, \dots, p_x$
        \item[] $s_{j+p_x} = \hat{x}_{k, k-1} - \sqrt{pd +\kappa} A_j, \quad j=1, \dots, p_x$
        \item[] $ \hat{\mu}_k = \sum_{j=0}^{2p_x} w_j \mu(s_j, \beta) $
        \item[] $ R_k = \hat{\mu}_k \hspace{0.01 \textwidth} I_{p_y}$
        \item[] $ S_k = \sum_{j=0}^{2p_x} w_j (\mu(s_j, \beta) - \hat{\mu}_k)(\mu(s_j, \beta) - \hat{\mu}_k)^{'} + R_k$
        \item[] $ C_k = \sum_{j=0}^{2p_x} w_j (s_j - \hat{x}_{k|k-1})(\mu(s_j, \beta) - \hat{\mu}_k)^{'} $
        \item[] $ K_k = C_k S_k^{-1}$
    \end{itemize}
    \item \textit{Filter update step}
\end{enumerate}    
\EndFor
\end{algorithmic}
\end{algorithm}

\subsection{Computational aspects}

The $p^2 \times p^2$ matrix inversion in (\ref{filteringMatrix}) represents a  computational bottleneck in many Kalman filter applications. However there are cases where the dimension of the latent process is much smaller than the observed process dimension. The Sherman-Morrison-Woodbury identity can be employed 
\begin{equation}
\notag
\left( R_k + H_k V_{k|k-1} H_k'\right) ^{-1} = R_k^{-1} - R_k^{-1} H_k   (V_{k|k-1}^{-1} + H_k' V_{k|k-1} H_k )^{-1} H_k' R_k^{-1}
\end{equation}
and requires $p \times p$ matrices inversion only.
As the latent space employed by our model has a cheap $p$-dimensional  representation our scenario is particularly appealing for the application of the Sherman-Morrison-Woodbury identity. The identity is closely related the Information Filter, see the Appendix \ref{D}, which usage is equivalent.
The overall computational cost of the algorithm is therefore dominated by the inversion of a $p \times p$ matrix \citep{mandel2006efficient}. 

\subsection{Model selection}
\begin{figure}[t]
    \begin{subfigure}{.5\textwidth}
      \centering
      \includegraphics[width=.8\linewidth]{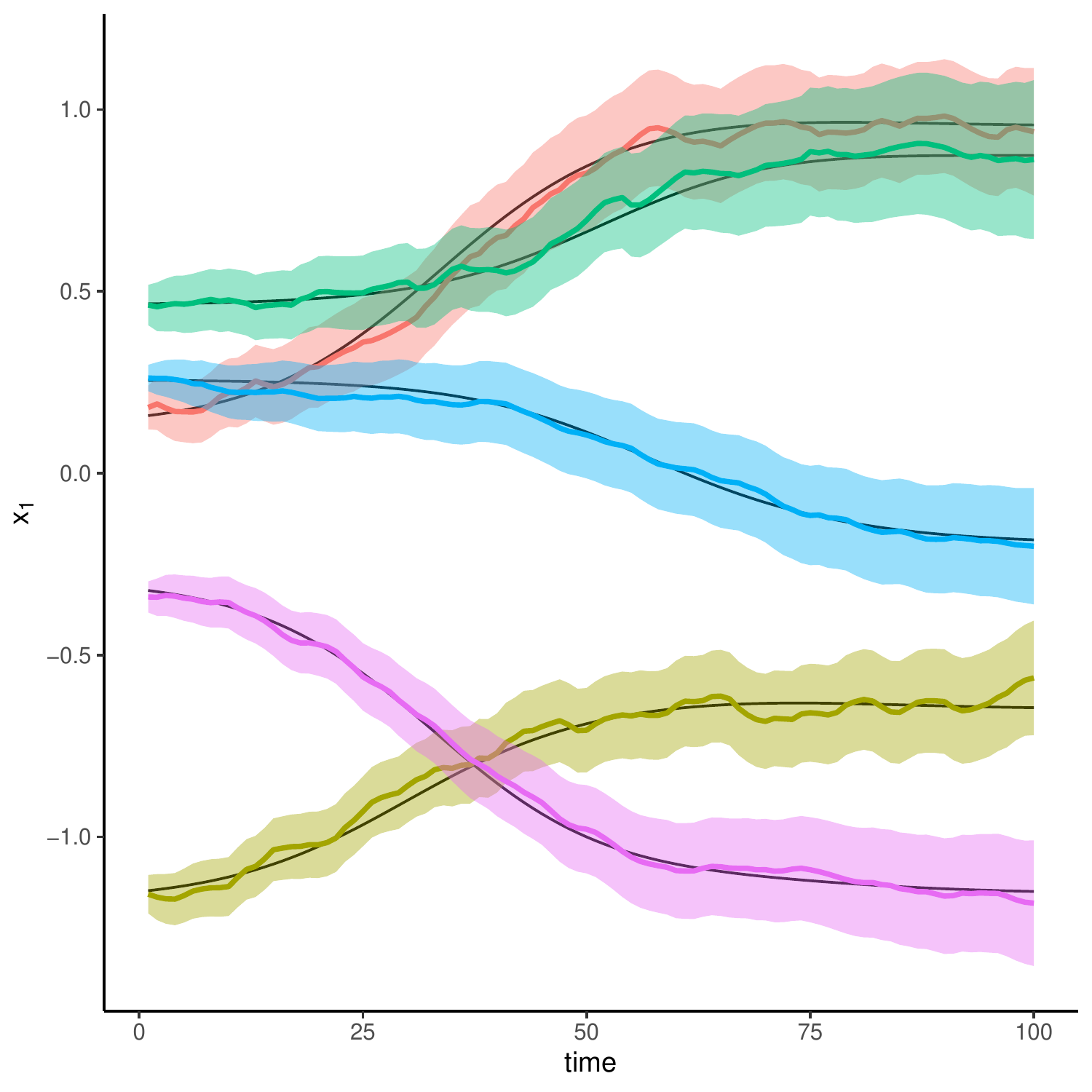}
    \end{subfigure}%
    \begin{subfigure}{.5\textwidth}
      \centering
      \includegraphics[width=.8\linewidth]{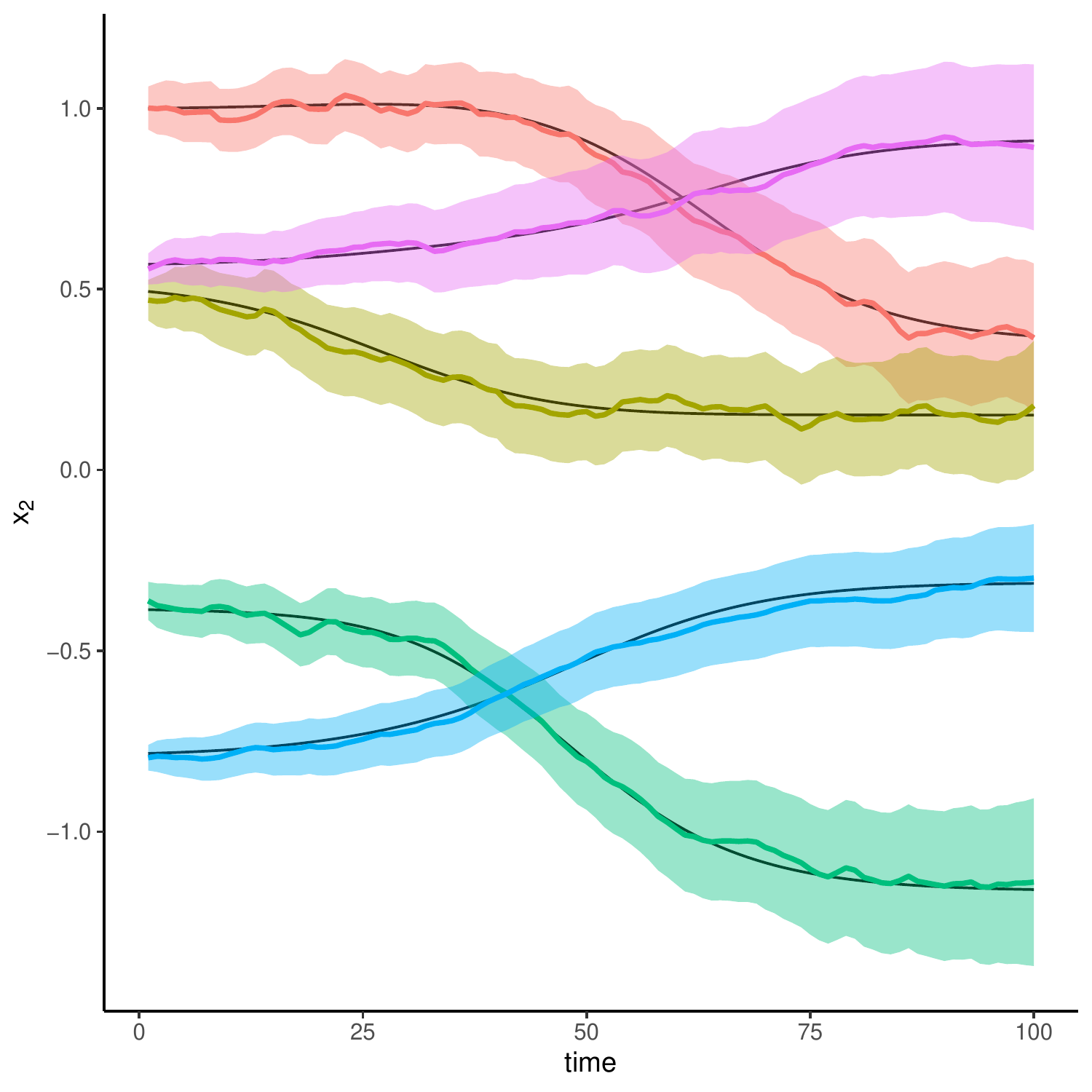}
    \end{subfigure}
    	\caption{An example of the model fit on simulated data with 10 dynamic nodes. On the right we present some of the estimated locations $x_{k|k}$ and their variability bands $x_{k|k} \pm 1.96\sqrt{V_{k|k}}$. Such quantities are produced by the smoother, allowing a straightforward assessment of the model uncertainty. The black line represents the true locations that we are simulating from.}
    \label{fig:bands}
\end{figure}

The conditional distribution of the latent space $x$ conditioned to the observed process $y$ can be used for assessing the uncertainty about the latent process. Variability bands can be draw by using the quantiles of the distribution $x_{k|n} \sim N(\hat{x}_{k|n}, V_{k|n})$ and the user can visually check whether the dynamic locations are far from being a constant line, as shown in Figure \ref{fig:bands}.

\paragraph{Akaike Information Criterion.}
The dimension $d$ of the latent space can be selected by using some Information Criterion such as the cAIC
\begin{equation}
\notag
    \text{cAIC}= -2 \log f(y|\hat{\beta}, \hat{x}) + 2 \Phi
\end{equation}
where $\Phi$ is the effective degrees of freedom of the fixed and random latent part of the model. \cite{saefken2014unifying} present a unifying approach for calculating the conditional Akaike information in generalized linear models that can be used in this context. 
This allows us to select the latent space dimension $d$ that minimize the conditional Akaike criterion. The cAIC is also used for making selection over the two filters, EKF and UKF, or to choose between different $\Sigma$ structures, e.g. a diagonal matrix with either the same or different variance parameters. In the same way we use the cAIC to choose a static or a dynamic model. The static model, where all the locations are fixed in time, can be obtained with a modification of our algorithm. The static model can be viewed as a dynamic model with one single time interval, obtained by grouping together all the time intervals. The filtering procedure is reduced to the update of the locations at the starting point and at the single interval, with the convergence $\hat{\Sigma} \rightarrow 0$. 


\paragraph{Goodness-of-fit.}
We can assess the model goodness-of-fit in the same way as done in multivariate generalized linear models. Residuals plots can be useful for spotting violations of the assumptions, e.g., the latent space assumption, the family and thus the correct variance function. Although it is possible to inspect all $p(p-1)$ fits on the counts $y_k$, we recommend a cheaper way. Residuals can be inspected by plotting the sequence of locations $x_{k|n}$ where the links are colored differently according to the studentized residual. We can choose red links for large residuals and green for the small ones, with all the shades in the middle. In case the variance function is misspecified we expect to observe more red links for closer nodes. In case the latent space assumption is violated we expect to see red links evenly spread over the network.

\section{Simulation study}
\label{sec:applied}

\begin{figure}[t]
    \begin{subfigure}{.5\textwidth}
      \centering
      \includegraphics[width=.8\linewidth]{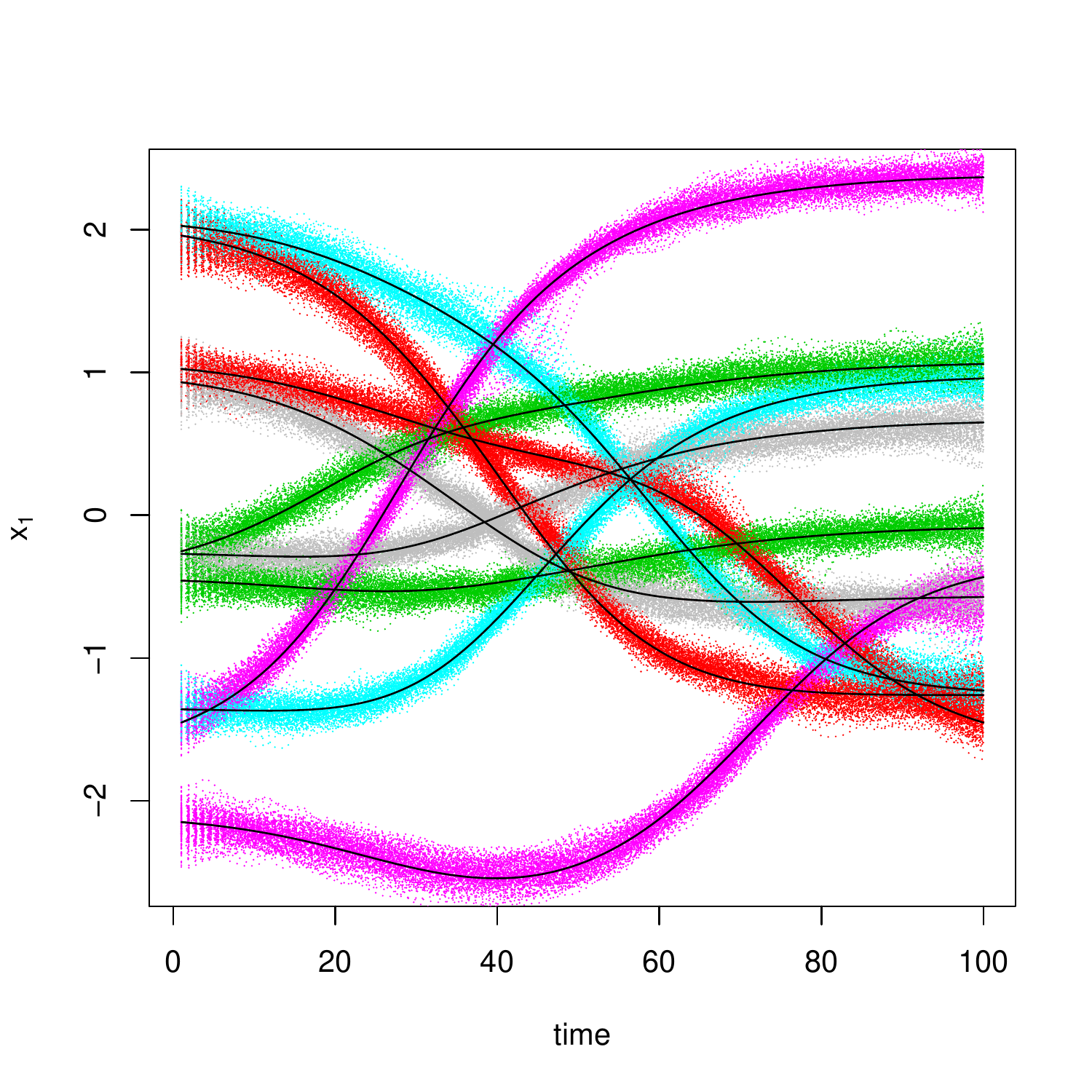}
    \end{subfigure}%
    \begin{subfigure}{.5\textwidth}
      \centering
      \includegraphics[width=.8\linewidth]{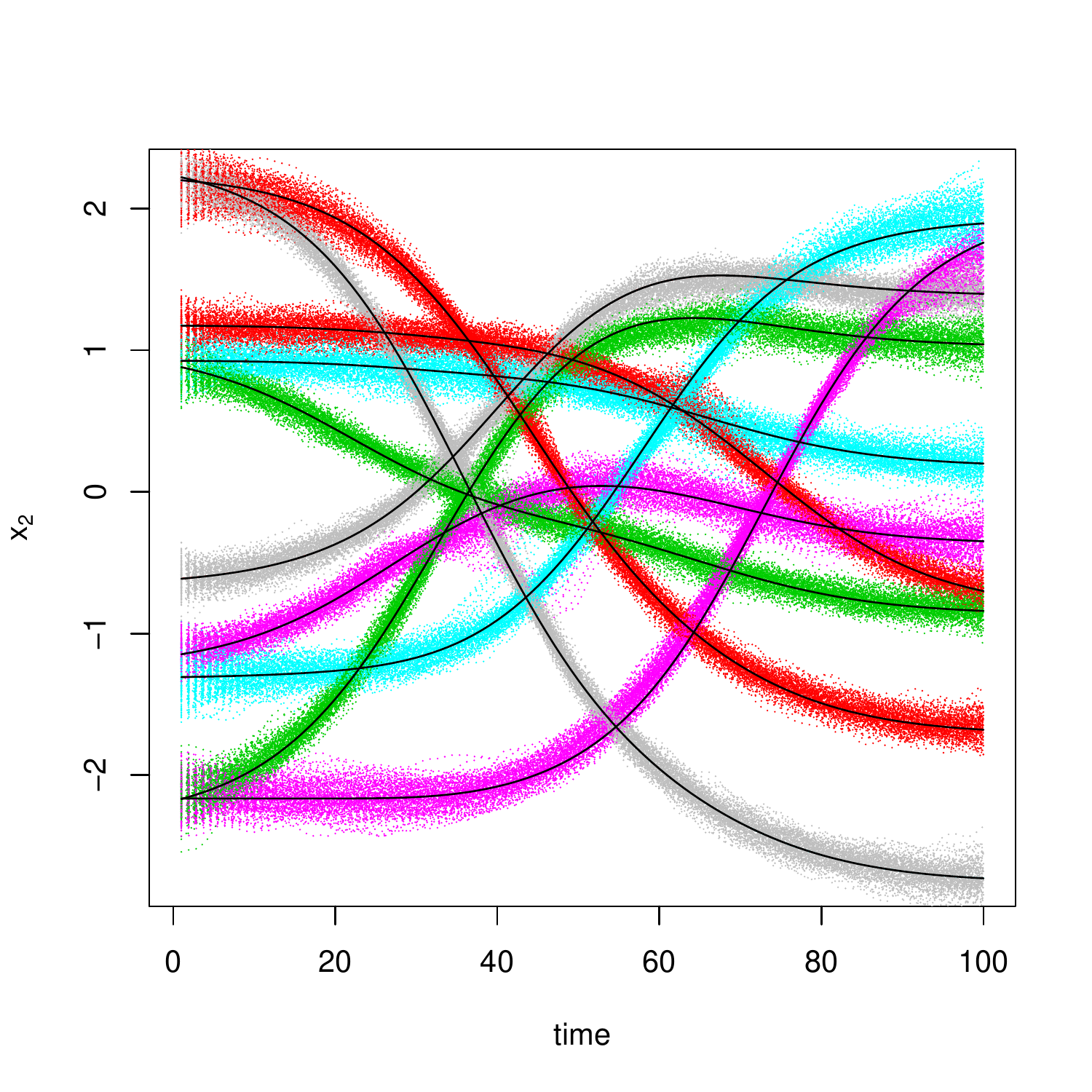}
    \end{subfigure}
    	\caption{Model fit on 200 simulated datasets. The figure shows that the estimated latent locations are centered at their true values with relatively high precision. Black lines represent the true locations in time. Colored lines represent nodes trajectories estimated by the model for each simulation. We consider $p=10, n=100, d=2$ and $x_1, x_2$ are respectively the first and the second dimension. }
    \label{fig:fitSim}
\end{figure}

In order to assess the method performance we carry out a simulation study.
We specify logistic functions for the latent location trajectories $x_k$,  rescaling and shifting these functions in different ways. The link counts are generated from a Poisson distribution with $\log(\mu_{ij}(x_k)) = \alpha - \|x_i(k)-x_j(k)\|_2^2$.
In Figure \ref{fig:fitSim} is shown a possible set of locations, the black lines. We simulated the observed $Y$ process 200 times from these trajectories. The colored lines are the 200  trajectories  estimated by the EM-EKF. We simulated with $p=10$ nodes,  $n=100$ intervals and $d=2$ dimensions.

The study that we carried on consists of a set of simulations that investigate the model behavior in different scenarios. 
We consider the model with $p=10, n=100, d=2$ and we vary the number of nodes, intervals and dimension. We also propose some challenges to the model such as the mispecification of the distribution family, high clustering or sparsity behavior. We also report the static model performances as a baseline for comparison. We use the out-of-fold Kullback Leibler divergence as performance measure
\begin{equation}
\notag
\begin{split}
   KL(\hat{x}, x_{\text{true}}) &= \mathbb{E}_y \left[ \log p(y|x_{\text{true}}) - \log p(y|\hat{x}) \right]   \\
& \approx \frac{\sum \log p(y_{\text{new}}|x_{\text{true}}) - \log p(y_{\text{new}}|\hat{x})}{n p(p-1)/2}
\end{split}
\end{equation}
where $y_{\text{new}}$ denotes an additional sample that is generated from $x_{\text{true}}$. The Kulback Leibler is a performance measure based on the distance matrix, which is invariant to rotations and translations of the locations.

\begin{figure}[t]
    \begin{subfigure}{.5\textwidth}
      \centering
      \includegraphics[width=.8\linewidth]{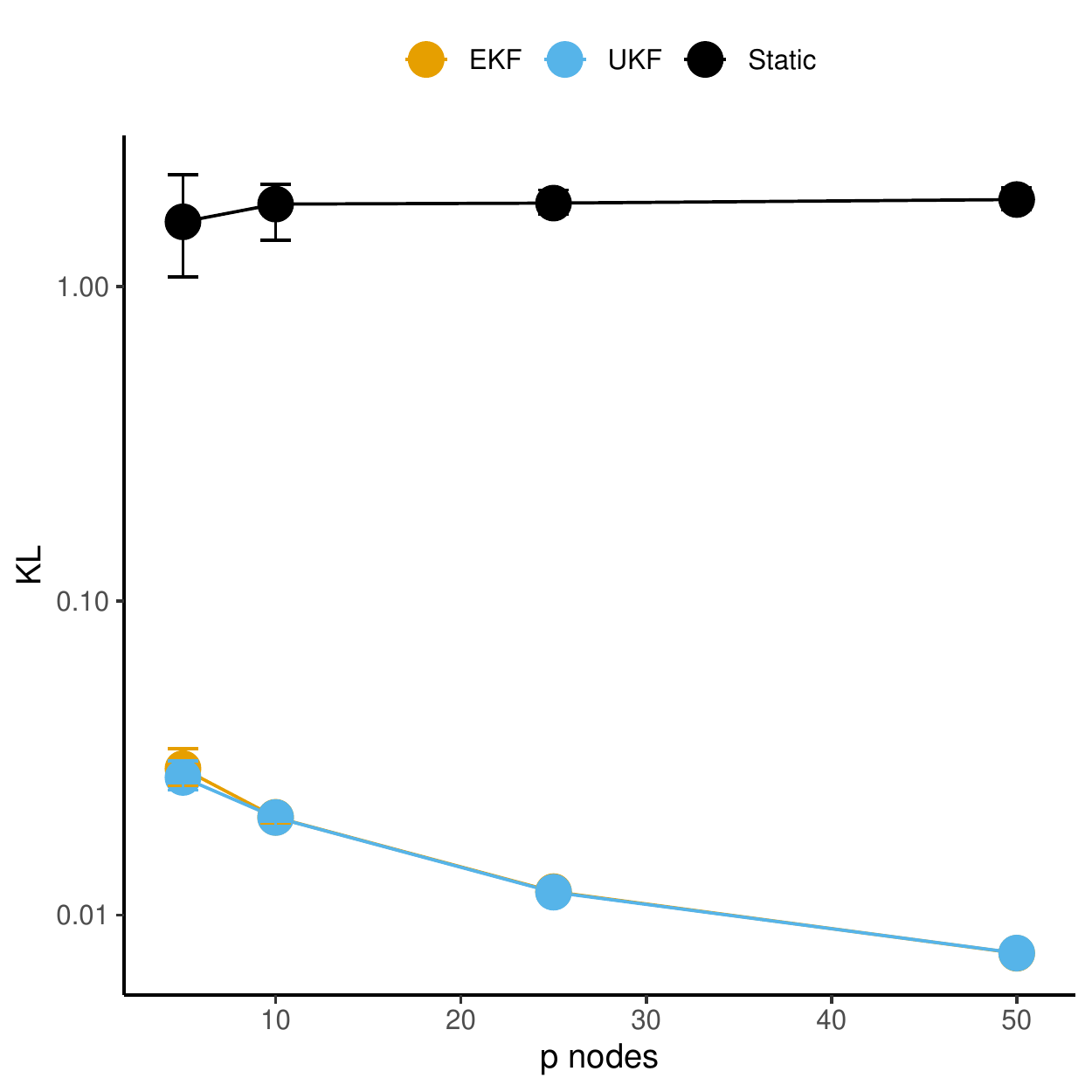}\\
      (a)
    \end{subfigure}%
    \begin{subfigure}{.5\textwidth}
      \centering
      \includegraphics[width=.8\linewidth]{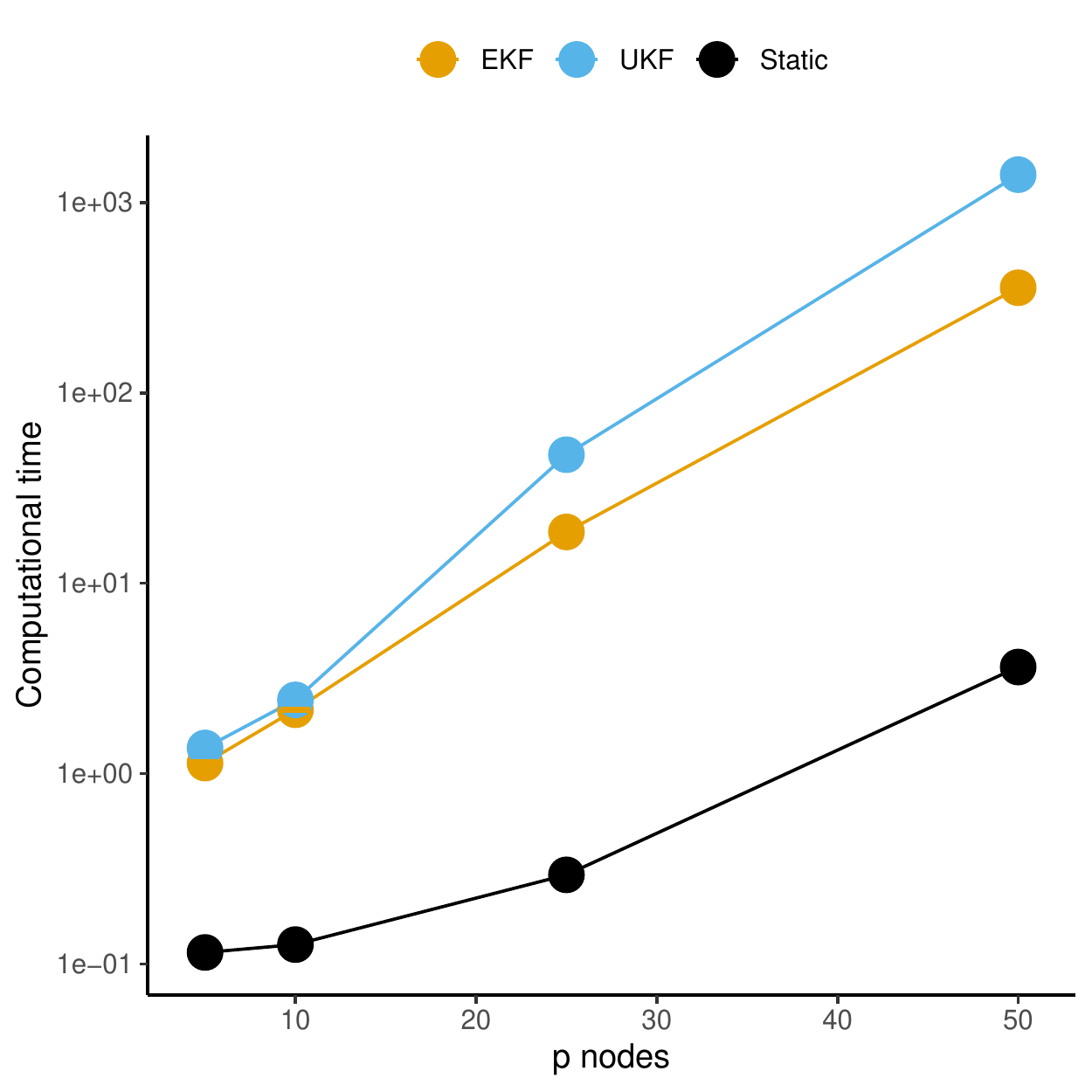}\\
      (b)
    \end{subfigure}
    	\caption{a. Kullback-Leibler measure shows that whereas the static model shows a stable misfit to the dynamic latent model, the EKF and UKF both improve performance with additional number of nodes $p$; b. Computational time grows markedly in the number of nodes $p$.}
    \label{fig:p}
\end{figure}

\begin{figure}[t]
    \begin{subfigure}{.5\textwidth}
      \centering
      \includegraphics[width=.8\linewidth]{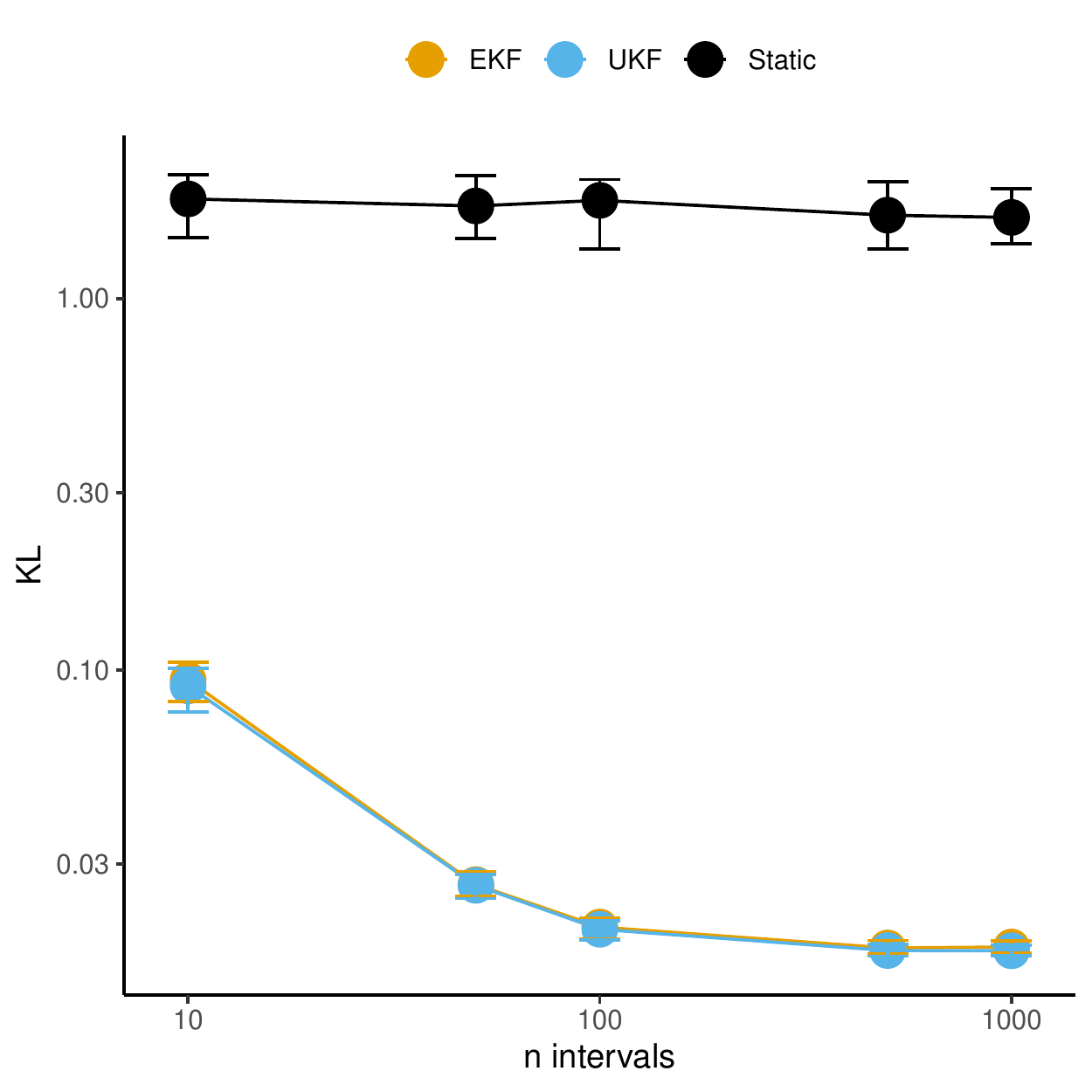}\\
      (a)
    \end{subfigure}%
    \begin{subfigure}{.5\textwidth}
      \centering
      \includegraphics[width=.8\linewidth]{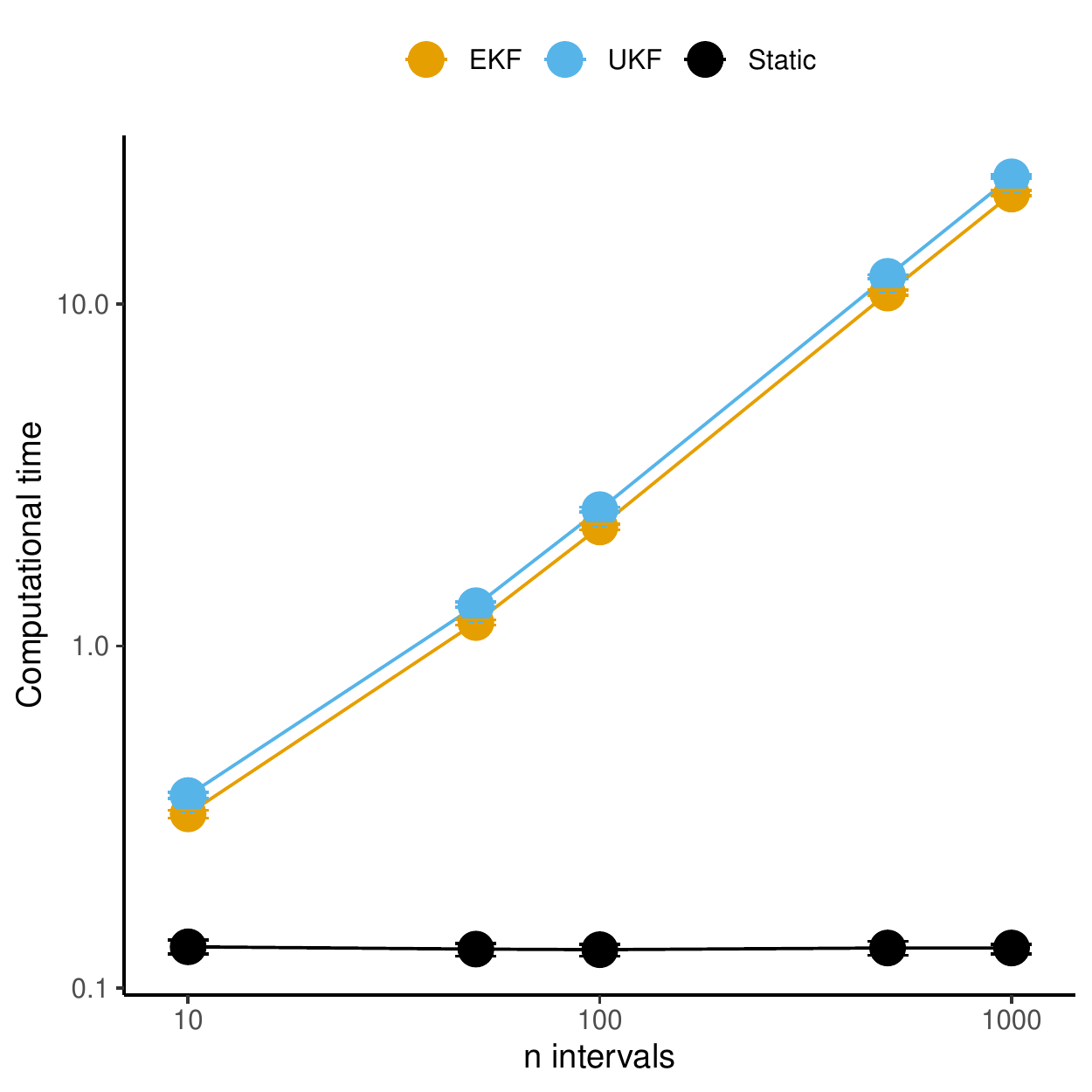}\\
      (b)
    \end{subfigure}
    	\caption{a. With increasing number of time points $n$ the Kullback-Leibler fit improves similarly for UKF and EKF, whereas the static model fit stays unchanged; b. the computational time grows linearly in $n$ for the UKF and EKF.}
    \label{fig:n}
\end{figure}


\paragraph{Varying the number of nodes p.}
Figure \ref{fig:p} shows the results of varying the number of nodes $p=5,10,25,50$. EKF and UKF have almost the same performance that improves as $p$ increases, as a consequence to the increment of information to our model. The dynamic latent space clearly outperforms the static model, whose KL fit remains stable with varying $p$.

\paragraph{Varying the number of intervals n.}
Figure \ref{fig:n} shows the results of varying the number of observed time intervals $n=10,50,100,1000$. For the dynamic models there is a strong performance improvement for low $n$, reaching a plateau beyond $n=100$ where adding other intervals does not have an important contribution to the KL. For $n=10$ we show that even for low number of intervals the dynamic model provides a better result than the static model. 

\paragraph{Varying the latent dimension d.}
We did  notice a slight decrease in the performance  when increase the latent dimension. This can find a possible explanation in the number of observations $np(p-1)$, which increase as we increase $p$ and $n$. The latent dimension $d$ gives no contribution to the number of observations and hence we observe no real difference in the performances.

\paragraph{Computational costs.}
Figure \ref{fig:n} shows that the computational cost grows approximately linearly with $n$, as the filter replicates the same matrix operations $n$ times. Differently to $n$  the computational costs in Figure \ref{fig:p} grow non-linearly with the number of nodes $p$, \cite{mandel2006efficient}. Similarly to the results in the performances, varying $d$ does not make a substantial difference in the computational costs.        

\begin{figure}[t]
    \begin{subfigure}[t]{.5\textwidth}
      \centering
      \includegraphics[width=.8\linewidth]{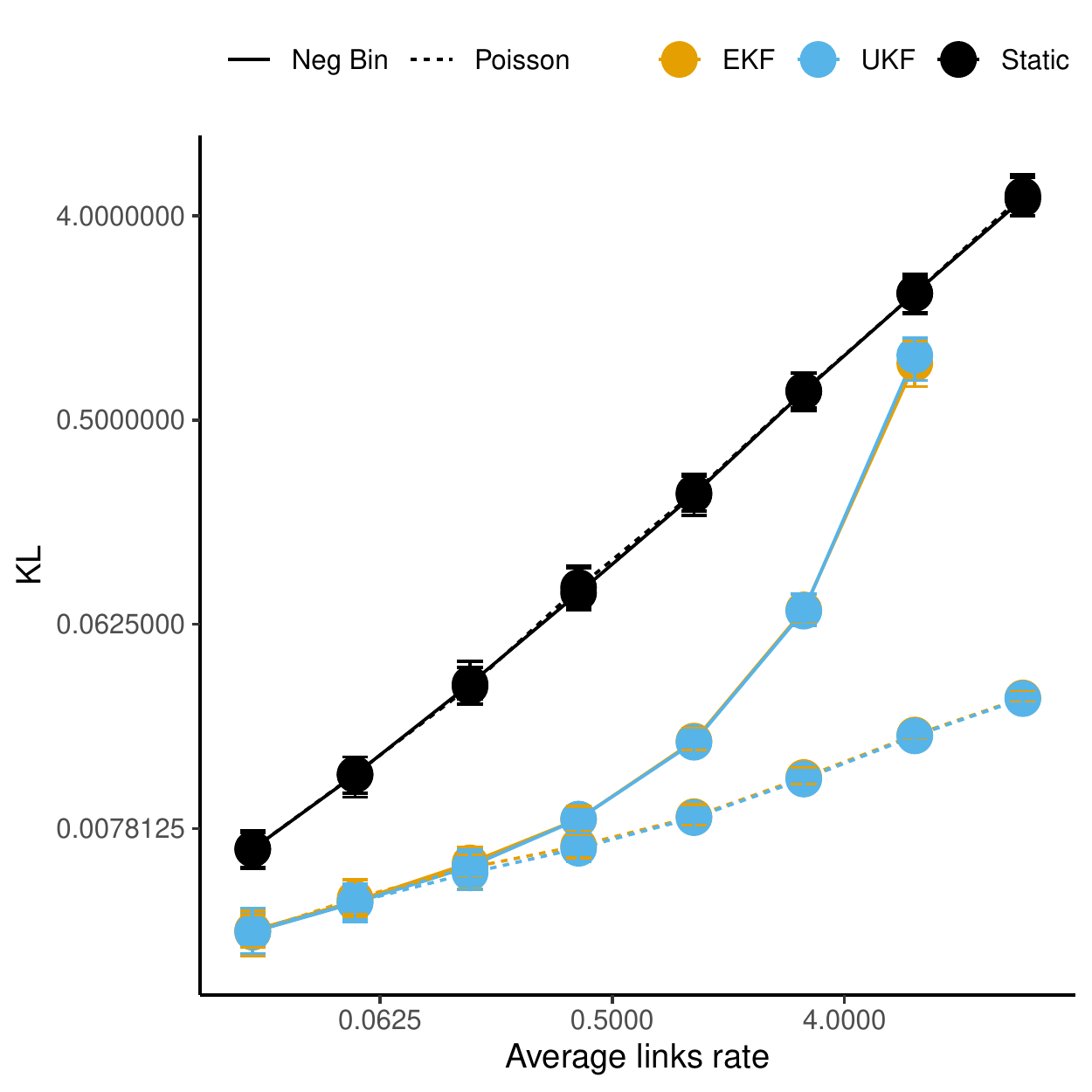}
    \end{subfigure}\hfill
    \begin{subfigure}[t]{.5\textwidth}
      \centering
      \includegraphics[width=.8\linewidth]{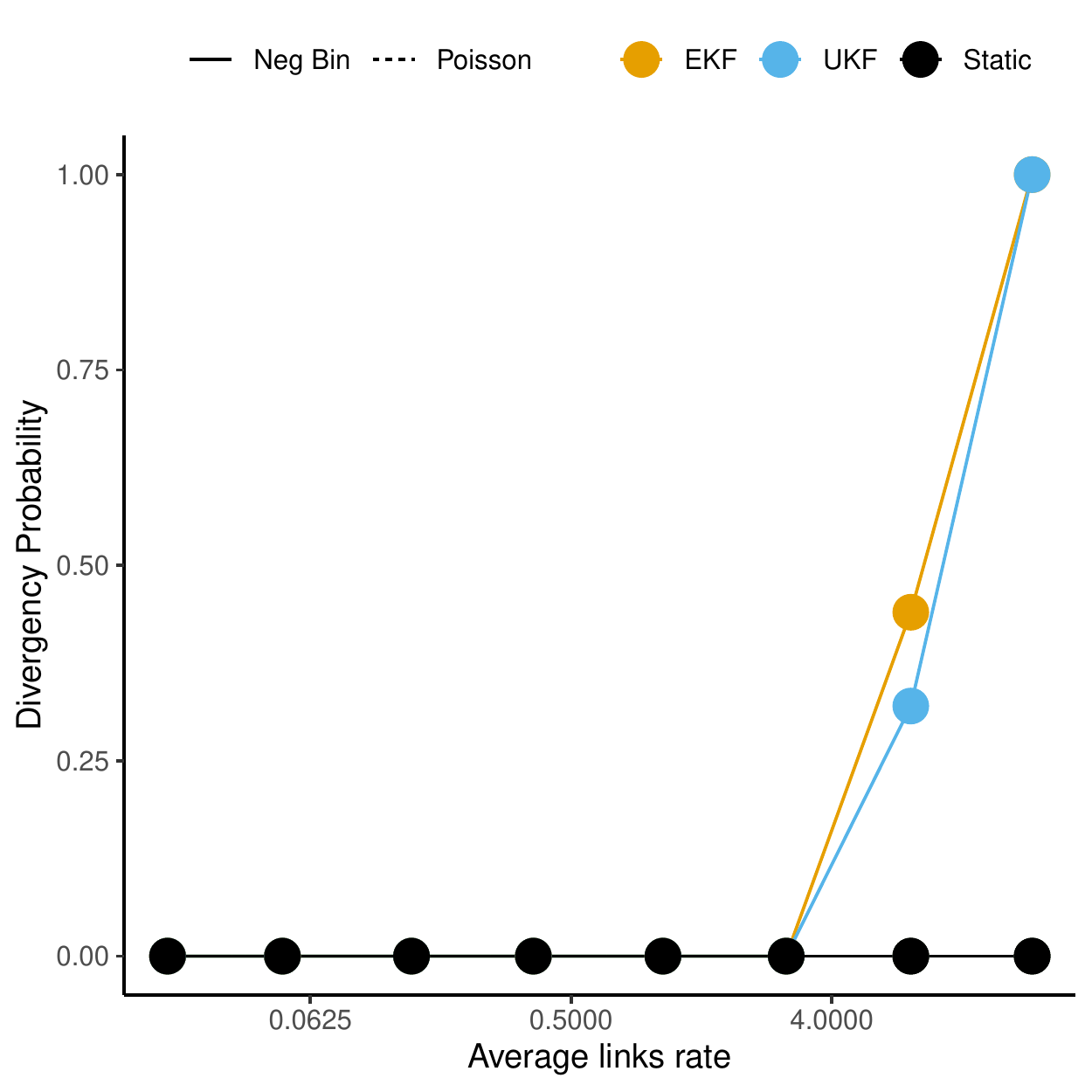}
    \end{subfigure}
    	\caption{Overdispersion vs correct family specification performances varying the rate of links in the network. The divergence frequency  suggests the level of overdispersion for which the model cannot retrieve the signal in the data.}
    \label{fig:sparse}
\end{figure}

\paragraph{Effect of overdispersion.}
In Figure \ref{fig:sparse} we investigate the model behavior under overdispersion. We simulate the data from a Negative Binomial  with mean $\mu_{ij}(x_k)$ and a quadratic variance function $\mu_{ij}(x_k)+\mu_{ij}(x_k)^2$ and compare it to data simulated from a Poisson distribution. We study the performance of our Poisson model under different ranges of rate $\mu_{ij}(x_k)$.
For low rates the Negative Binomial variance is almost the same as that of the Poisson, and here we observe the same performances over the two settings.  For high rates the fit on Negative Binomial counts get worse and is comparable to that of the static model. For the highest rate the signal-to-noise ratio in the data is so low that the model diverges in all the simulations. In these cases the solution is to change the distribution specification and fit it with the right variance function.

The average link rate is related to the sparsity in the observed counts $y$. Figure \ref{fig:sparse} shows that the model still work even in high sparsity settings without divergence problems. This allows  the user to freely specify a high number of intervals $n$ for the analysis.


\begin{figure}[t]
    \begin{subfigure}[t]{.5\textwidth}
      \centering
      \includegraphics[width=.8\linewidth]{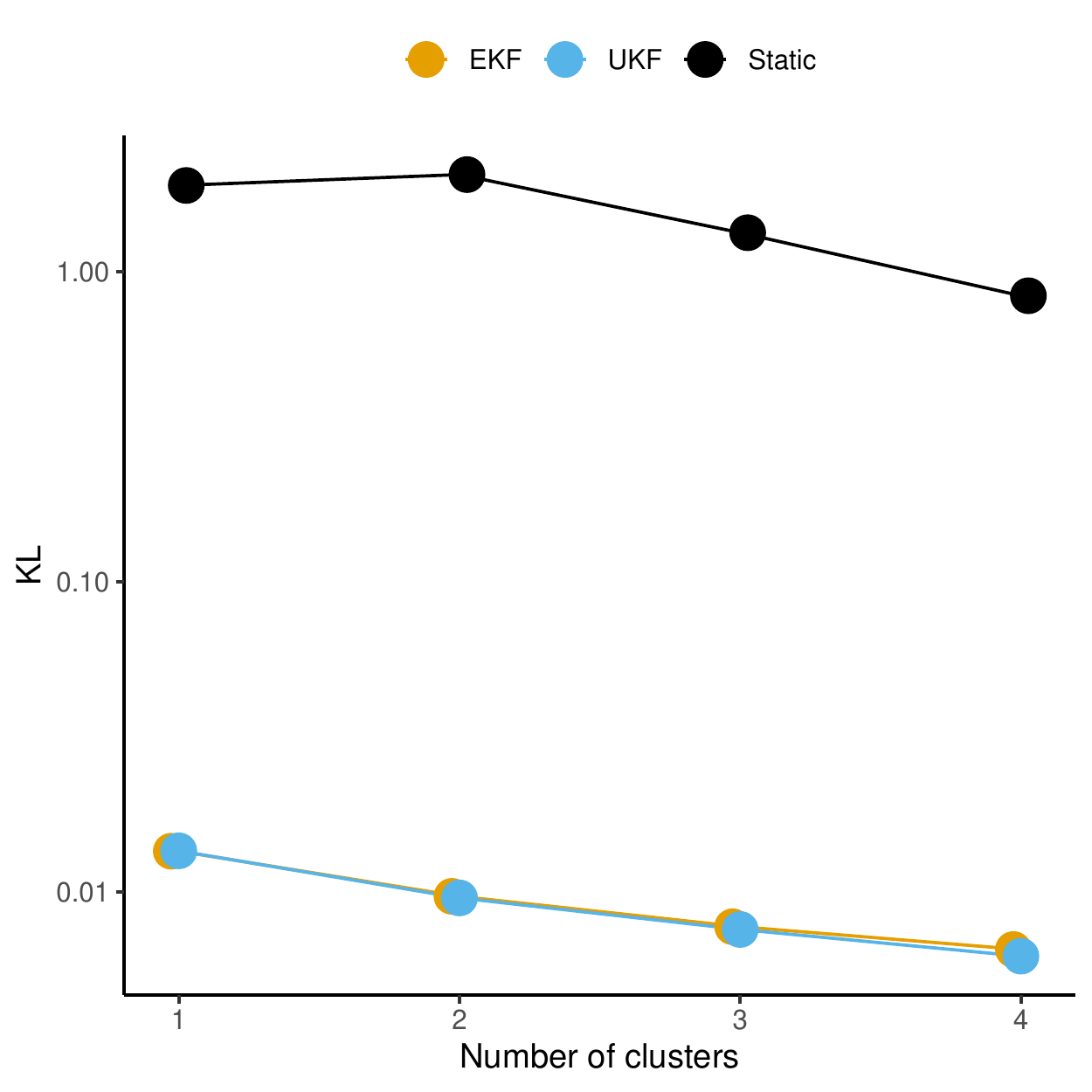}
    \end{subfigure}\hfill
    \begin{subfigure}[t]{.5\textwidth}
      \centering
      \includegraphics[width=.8\linewidth]{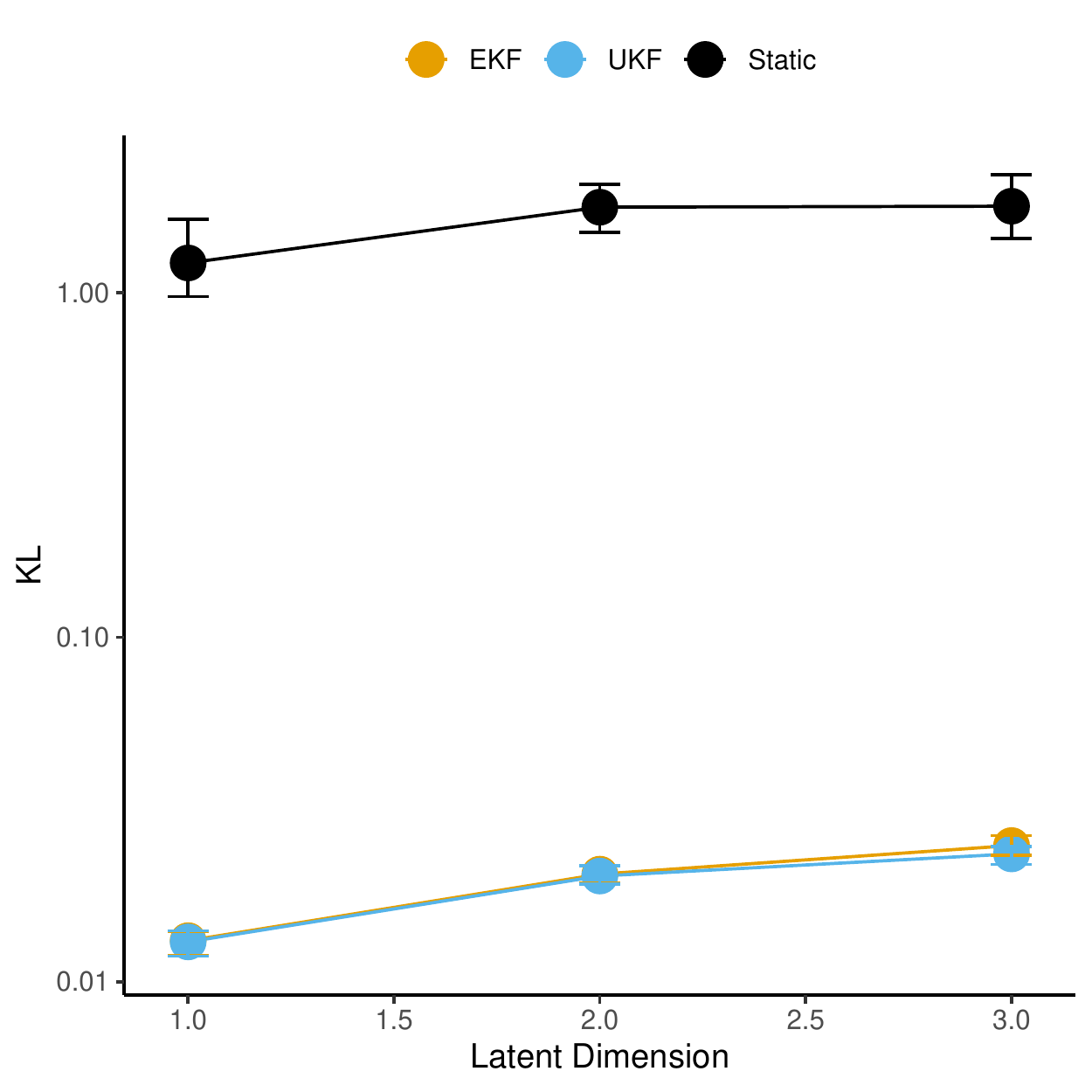}
    \end{subfigure}
    	\caption{KL measure by varying number of clusters in the simulated data or increasing the latent dimension. The model fit does not deteriorate with a higher dimension $d$ and does not change substantially when we have clusters formed in the latent space. The latter can be seen as a mixed sparsity scenario.}
    \label{fig:cluster}
\end{figure}

\paragraph{Considerations on identifiability.}
The latent formulation is identifiable in the relative distances but unidentifiable in the locations \citep{hoff2002latent}: infinite combinations of rotations and translations have the same distances and therefore the same likelihood. This implies the non-identifiability of $\Sigma$, as the coordinate system rotates.
Each update of the filter and smoother may involve a certain shift and rotation in the next location configuration. As a result when we update the starting points $x_{0|0}$ for the next EM iteration they may be shifted and rotated, with related rotation for $\Sigma$. These movements become stable as the starting points $x_{0|0}$ converge. In case identifiability is required in the analysis the user can specify $\Sigma$ spherical or spherical within each node, obtaining $\Sigma$  unaffected by rotations.

\paragraph{Considerations on filter divergence.}
A practical aspect that most Kalman Filter users deal with when working on real data is the divergence problem. Many factors can influence the divergence tendency such as a wrong variance function in $R_k$, poor approximation of non-linearity, inappropriate initial choice $\beta$, abrupt changes in link rates, too large variances $V_{0|0}$ and $\Sigma$. In those case $R_k$ is problematic and might then be approximated by $R_{k-1}$. In case of bad starting points $x_0$ the update of locations might have abrupt changes because in a non-convex likelihood optimization locations jump to find a more stable configuration.  

Fine-tuning parameters and starting points can make a difference, when divergence occurs. Problematic $R_k$ can be solved by taking more update steps on the same time point \citep{fahrmeir1992posterior}. Inflating $R_k$ solves overdispersion problems, although inferring the correct variance function of the data might take some extra effort. Sufficiently good $x_{0|0}$ points can be calculated via Multidimensional Scaling or reversing the time dimension and run the Kalman Filter backward.
Furthermore, we recommend starting the EM from the static model, thus $\Sigma$ low, and then expand it slowly toward the maximum likelihood point, as starting with a high $\Sigma$ and $V_{0|0}$ may overfit the data. In most pathological cases the model diverges before reaching the maximum likelihood point and a profile maximum likelihood estimate will be the best alternative. Another delicate aspect is the rate function choice. The function $e^{- \|x_i(k)-x_j(k)\|_2^2}$ is appealing because is differentiable. However it can be more unstable than other non-differentiable functions that exhibit a weaker non-linearity.
Every choice brings different complications and there is not an optimal choice for all scenarios.

\section{Dynamics of patent citation patterns}
\label{sec:patent}

The patent citation process presents some peculiar characteristics: patents are continuously added to the system and the citations happen in the moment of the patent creation only. A patent can cite only patents that are previously added and not the ones that are added in the future. 
In this analysis we group all these patents by the same  \href{https://ipcpub.wipo.int/?notion=scheme&version=20190101&symbol=none&menulang=en&lang=en&viewmode=f&fipcpc=no&showdeleted=yes&indexes=no&headings=yes&notes=yes&direction=o2n&initial=A&cwid=none&tree=no&searchmode=smart}{ICL class} and we use these fields as the unit of our analysis. Since there is a continuous exchange of citations between the fields, the resulting process can be regarded as a point process.  
The classification is the following
\begin{itemize}
    \item[A]: Human necessities.
    \item[B]: Performing operations; Transporting.
    \item[C]: Chemistry; Metallurgy.
    \item[D]: Textiles; Papers.
    \item[E]: Fixed constructions.
    \item[F]: Mechanical Engineering; Lightning; Heating; Weapons; Blasting.
    \item[G]: Physics.
    \item[H]: Electricity.
\end{itemize}
although other grouping schemes are possible, see \cite{younge2016patent}. The patent citation data are
available from \url{https://sites.google.com/site/patentdataproject/Home} 
and consists of 3.1 millions patents, 23.6 millions citations over the period 1967-2006, with collection intervals of 1 year length.  We consider the latent space model  
\begin{equation}
\label{eqn:rescale}
\begin{split}
&    Y_{ij}(k) \sim \mbox{Poi}(\mu_{ij}(x_k, \beta)) \\
&    \log(\mu_{ij}(x_k, \beta)) = \log(C_i(k)) + \alpha_0 -\|x_i(k)-x_j(k)\|_2^2  + \text{sender}_{i} + \text{receiver}_{j}
\end{split}\
\end{equation}
where $i$ and $j$ are two fields, $\alpha_0$ is an intercept and $\text{sender}_{i} $ and $\text{receiver}_{j}$ are respectively, the sender and receiver random effects. The citation rate is proportional to the number of patents added in a field within a year. If in a certain year there are no patents added in a field, the rate must be set to 0. We therefore specify an additional offset $C_i(k)$ that  account for the number of patents added in field $i$ at time $k$. The inclusion of $C_i(k)$ brings a different interpretation and hence we are modeling the citation rate per single patent in class $i$. We consider a bidimensional latent space for the sake of visual representation. 

\begin{figure}
      \centering
      \includegraphics[width=0.5\linewidth]{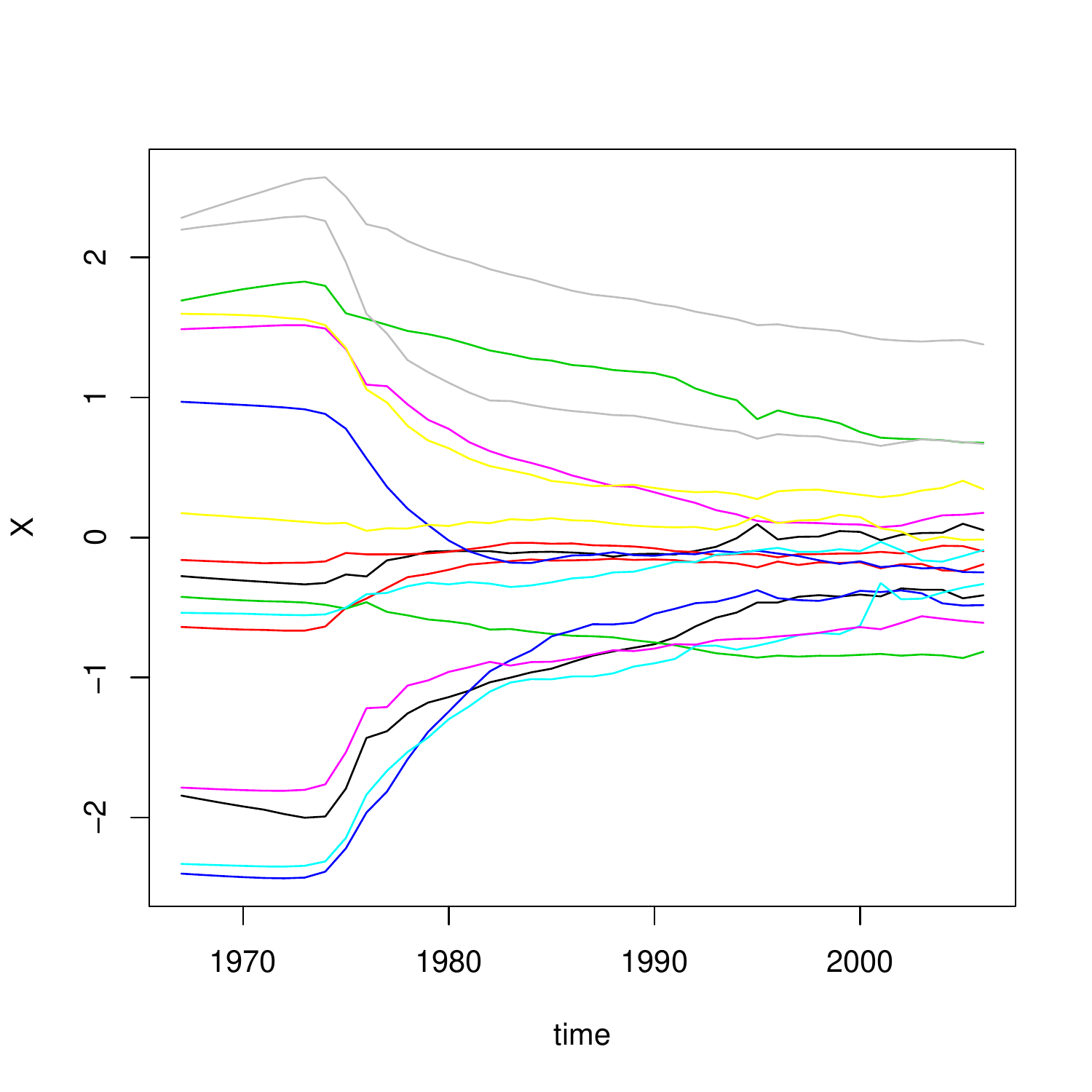}
      \caption{Changes in patent citation pattern, the two coordinates on the same plot. The first ten years show a static behavior in citations. After that point the fields start moving toward a closer form as the citations between fields intensify. }
    \label{fig:converge}
\end{figure}

\begin{figure}
    \begin{subfigure}[t]{0.5\textwidth}
      \centering
      \includegraphics[width=0.9\linewidth]{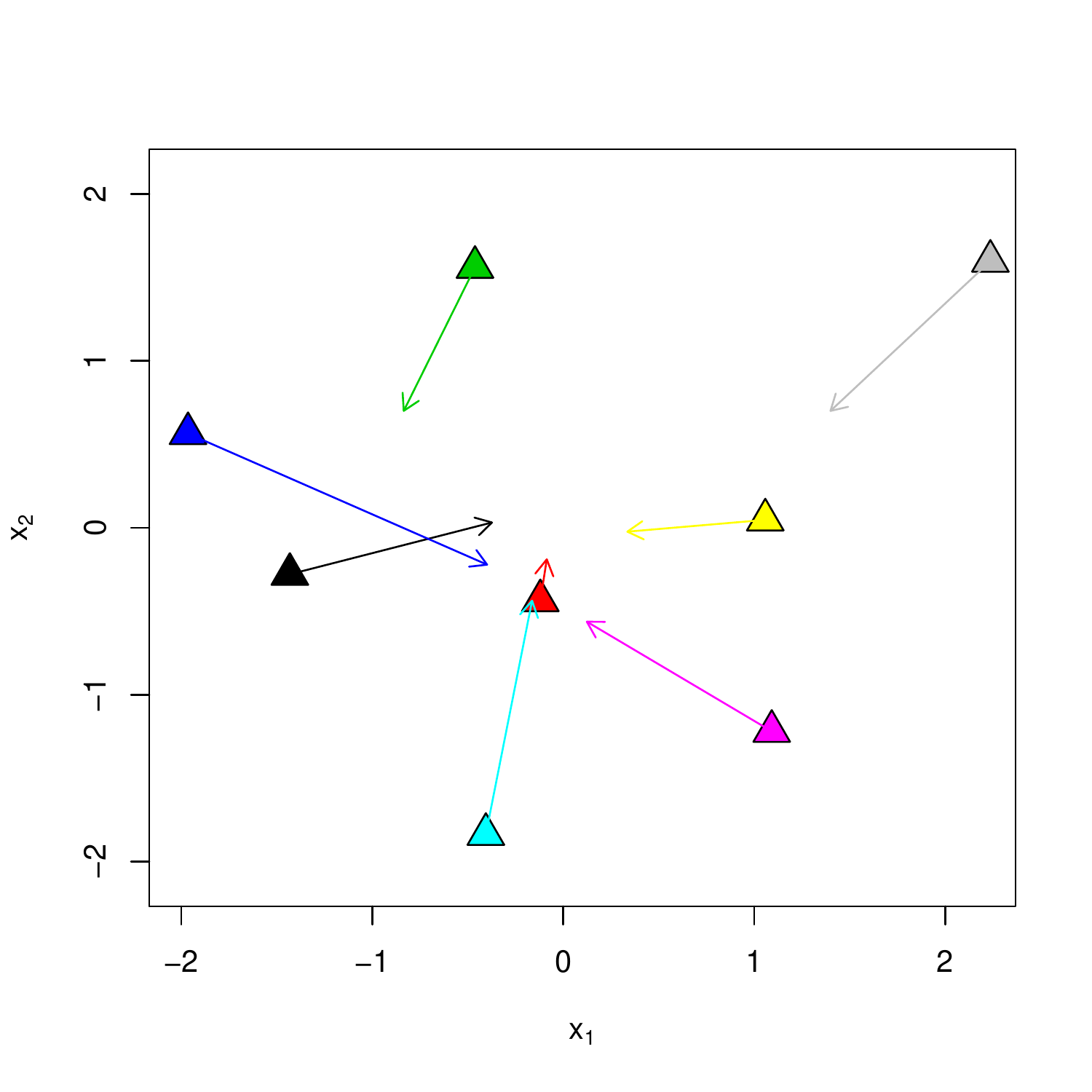}
      \caption{Changes in patent citation pattern. Interval years 1967-2006.}
    \end{subfigure}\hfill
    \begin{subfigure}[t]{0.5\textwidth}
      \centering
      \includegraphics[width=0.9\linewidth]{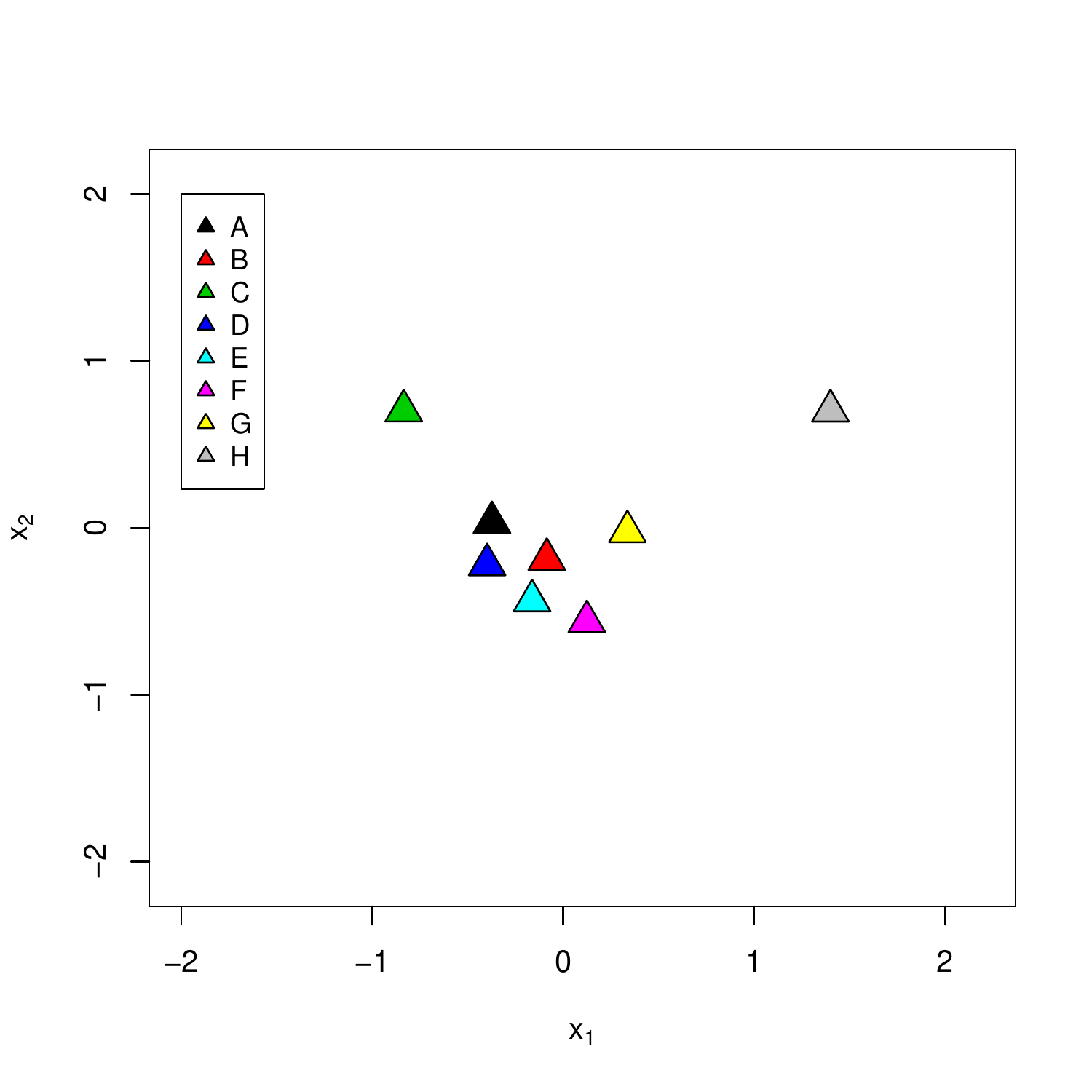}
      \caption{Final configuration.}
    \end{subfigure} 
    \vfill
    \begin{subfigure}[t]{0.5\textwidth}
      \centering
      \includegraphics[width=0.7\linewidth]{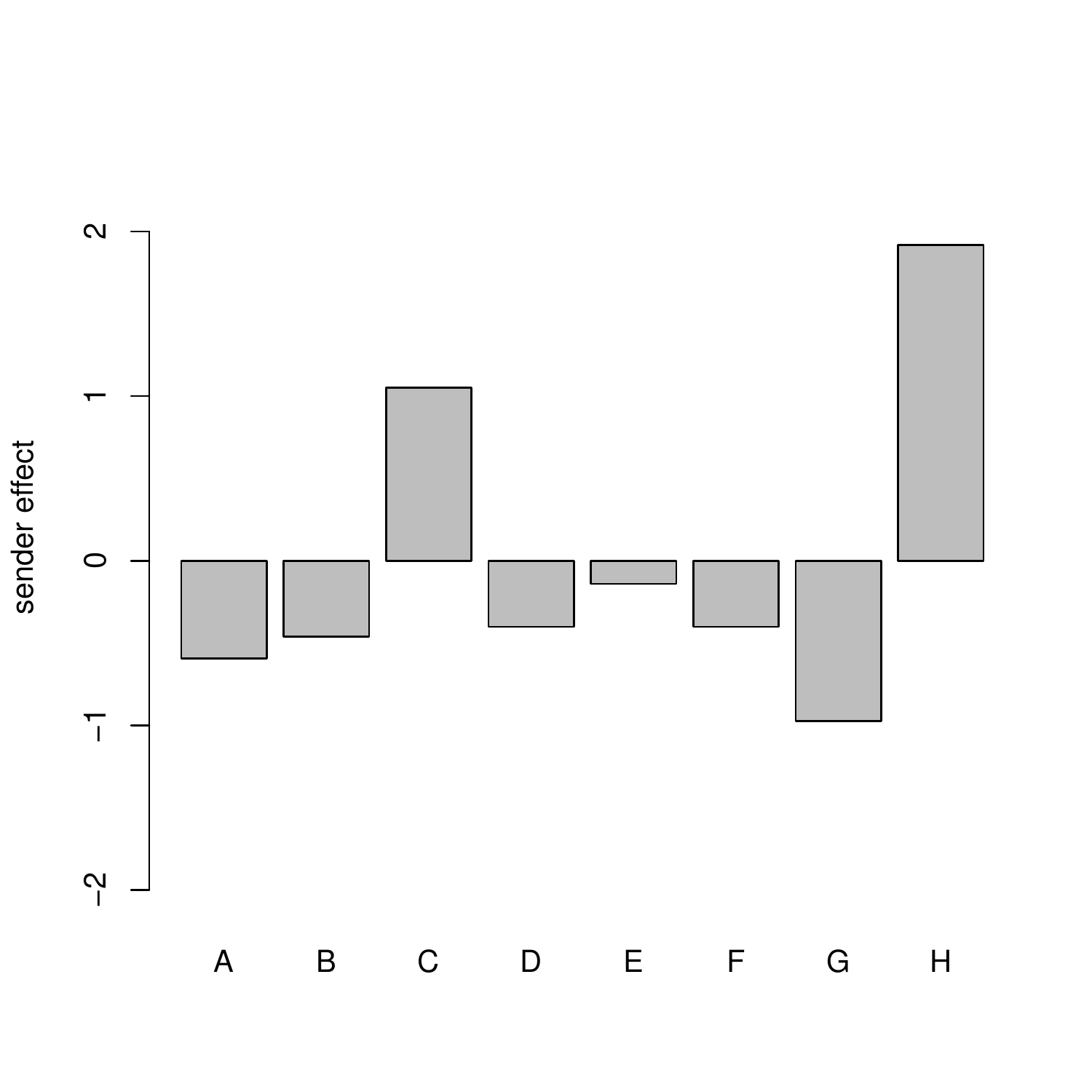}
      \caption{Sender effect}
    \end{subfigure}\hfill
    \begin{subfigure}[t]{0.5\textwidth}
      \centering
      \includegraphics[width=0.7\linewidth]{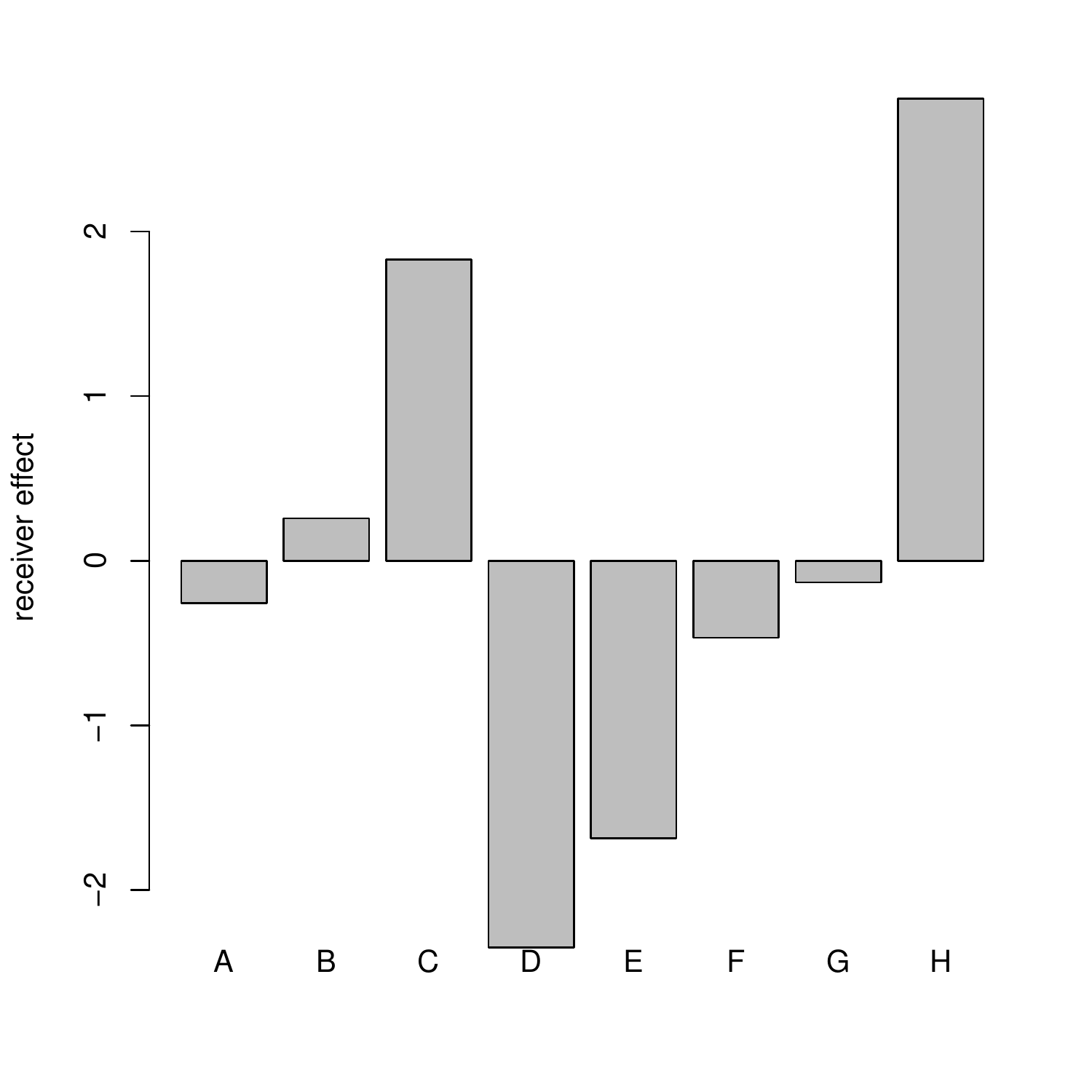}
      \caption{Receiver effect}
    \end{subfigure}
    \caption{Model inference on dynamic locations for the relational event model with sender and receiver effects. (a) shows a summary of the movement of the patent classes in the observed time interval.}
    \label{fig:patent1}
\end{figure}

\begin{figure}
    \begin{subfigure}[t]{0.5\textwidth}
      \centering
      \includegraphics[width=0.9\linewidth]{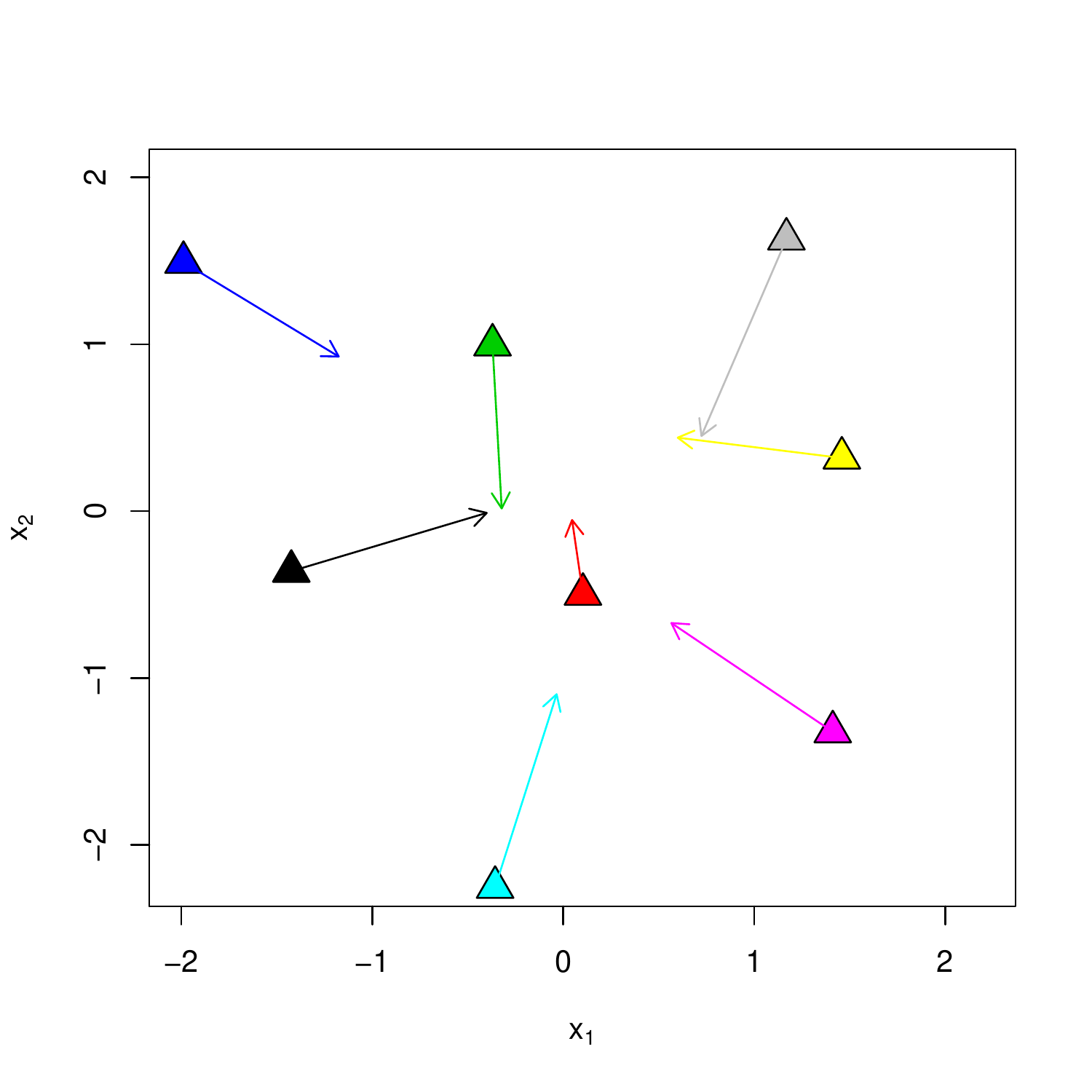}
      \caption{Changes in patent citation pattern. Interval years 1967-2006.}
    \end{subfigure} \hfill
    \begin{subfigure}[t]{0.5\textwidth}
      \centering
      \includegraphics[width=0.9\linewidth]{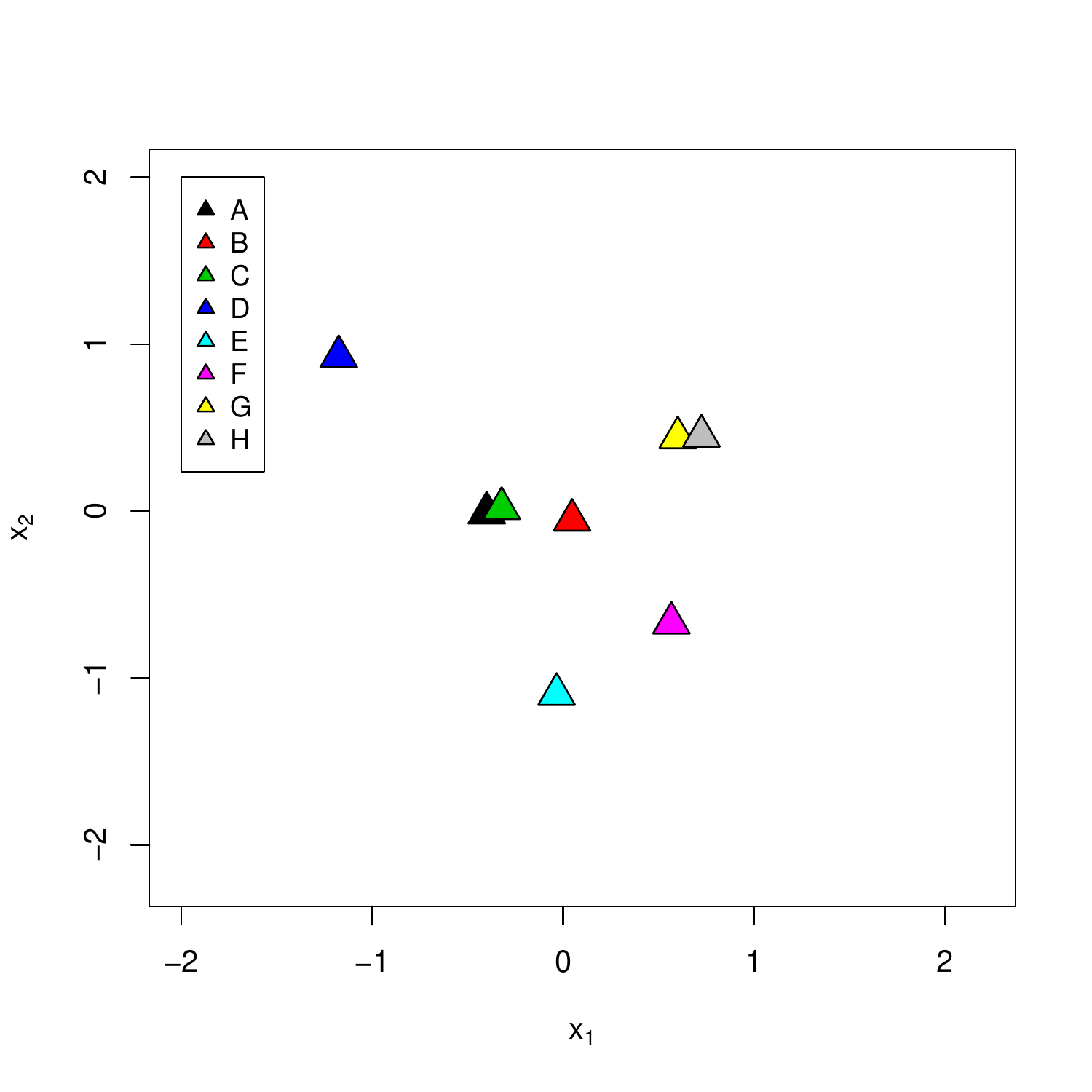}
      \caption{Final configuration.}
    \end{subfigure}  
    \caption{Model inference on dynamic locations for the relational event model without sender and receiver effects.}
    \label{fig:patent2}
\end{figure}

We fitted both the EM with EKF and UKF obtaining similar results, as anticipated by the simulation study. Figure \ref{fig:patent1} presents the estimated locations for the fields as well as sender and receiver effects. The legend letters match the mentioned classification of fields.

The sender and receiver effects can be interpreted as the asymmetry between fields citations that the latent space representation fails to capture.
Figure \ref{fig:patent1}(d) show how the Textile, Papers and Fixed constructions classes are very low receiver classes, meaning that they are cited below average. Figure \ref{fig:patent1}(c) shows that Physics patents a low tendendency to cite others.  The high sending and receiving tendencies of the Chemistry, Metallurgy and Electricity patents must be seen in the context of Figures \ref{fig:patent1}(a) and (b): the fact that we observe such huge effects jointly together with their distant location to the other patent classes might suggest some violation of the model assumptions. The two locations should be closer to the main cluster but there is not a latent configuration that makes a good fit. For comparison we fit the model without random sender and receiver effects: Figure \ref{fig:patent2}(b) shows that the distances of the Chemistry, Metallurgy and Electricity patent classes were inflated and that the random sender and receiver effects were indeed capturing the misrepresentation. The Physics patents comes now very close to Electricity, whereas the Chemistry and Metallurgy class overlaps with Human necessities. By looking back at the discrepancy between sender and receiver effects we see that Chemistry and Metallurgy patents have the tendency to receive more from Human necessities, whereas the Physics patents receive more citations from Electricity. In Figure \ref{fig:patent2}(b) Textile, Papers and Fixed constructions classes are pushed far away as the latent space accounts now for their negative receiver effect. 

Figure \ref{fig:converge} shows a peculiar behavior as locations are static in the initial 10 years. Patents can only cite back in time and therefore the first patents added in the system cannot cite patents submitted before the year 1967. The Figure suggests that around 1976 the patent citation process start behaving ``correctly'', i.e., that the database starts to include most cited patents. This seems reasonable as patents cite an average of 10 years back in time, with a mode that is significantly less than 10 years. 

In general we can observe that the exchange of citations between different fields increases trough time, ending with a large cluster including the majority of the ICL categories. The overall conclusion for this analysis on the Patents data is that there is an increment in the connectivity between different fields.  This suggests that most technology classes are becoming less dissimilar: there is an increasing heterogeneity within the fields, as they communicate with other technology fields, and thus a higher homogeneity between the fields.

\section{Conclusion}
In the last decade REMs have been used for describing the drivers of dynamic networks interactions. Traditional approaches focus on endogenous and exogenous drivers, which may not always be able to capture all heterogeneity in the data. Our aim has been to extend relational event modelling by letting their interactions depend on dynamic locations in a latent space.

Our estimation approach of the latent space relational event model combines several methods: the Expectation Maximization algorithm, Kalman filters and Generalized Additive Models. 
We consider the latent locations as missing states. The filter calculates their conditional expectation and the Generalized Additive Model performs the maximization: the two main ingredients for an EM algorithm. Kalman Filters are effective methods for estimating latent dynamic processes. Their simplicity and intuitive usage make them suitable for many problems, commonly  in engineering contexts. The filter relies on a sequence of linear operations and easily calculates the Expectation step, typically untractable for non-trivial cases. The Kalman filter dual interpretation in both the Bayesian and frequentist literature would also make an effective within-Gibbs implementation, instead of a within-EM implementation, possible.
The sequence of updates in the latent space makes the Kalman filter an effective tool for tracking the movements of the latent locations, as already proved in many applications. Our model formulation is very general and can encompass all the Generalized Additive Model features such as fixed effects, random effects and smoothly time-varying effects. 

The simulation results show that the model is accurate, computationally feasible and insightful under different scenarios. The patent citation analysis gives an interesting interpretation on innovation dynamics in the period 1967-2006 where many traditionally distinct patent classes show a marked convergence in a latent knowledge space.

\bibliographystyle{apalike}
\bibliography{references}
\newpage


\appendix
\section{Appendix}

In (\ref{eqn:ss}) $x_k$ and $y_k$ are vectors of length $p_x = pd$ and $p_y = p(p-1)$ or $p(p-1)/2$ in case of an undirected network respectively. These are the the $p\times d$ location matrix and $p\times p$ adjacency matrix that have been vectorized. At time $k$ we have  
\begin{equation}
	\notag
	\mu(x_k,  \beta) = 
	\begin{bmatrix}
		\mu_{1, 2}(x_k, \beta) \\
		\vdots \\
		\mu_{p-1, p}(x_k, \beta) 
	\end{bmatrix}
	,
	\hspace{0.05\textwidth}
	x_k = 
	\begin{bmatrix}
		x_1(k) \\
		\vdots \\
		x_p(k) 
	\end{bmatrix}
	,
	\hspace{0.05\textwidth}
	x_i(k) =
	\begin{bmatrix}
		x_{i1}(k) \\
		\vdots \\
		x_{id}(k) 
	\end{bmatrix}
	,
\end{equation}
where $x_{i}(k)$ is the $d$-dimensional location of node $i$.
The choice of using the euclidean distance is arbitrary and other distance measures can be selected. The dimension of the latent space is commonly chosen as $d=2$ or $3$ for sake of visual inspection, but more formal criteria can be used to select a proper dimension.

The matrix $H_k$ of the first derivatives  is structured as follows
\begin{equation}
\notag
H_k = \frac{\partial }{\partial x} \mu(x,\beta) \left. \right|_{\hat{x}_{k|k-1}} = 
\begin{bmatrix}
\frac{\partial }{\partial x} \mu_{1, 2}(x,\beta) \left. \right|_{\hat{x}_{k|k-1}}\\
\vdots \\
\frac{\partial }{\partial x} \mu_{i, j}(x,\beta) \left. \right|_{\hat{x}_{k|k-1}}\\
\vdots \\
\frac{\partial }{\partial x} \mu_{p-1, p}(x,\beta) \left. \right|_{\hat{x}_{k|k-1}}  
\end{bmatrix}
\end{equation}
$H_k$ is a $p_y \times p_x$ block matrix, where the row indexed by the interaction $(i,j)$ is composed of $d$-dimensional vectors $\frac{\partial }{\partial x_k} \mu_{i,j}(x, \beta)$ for $k= 1, \dots, p$ as follows
\begin{equation}
\notag
\frac{\partial }{\partial x} \mu_{i,j}(x,\beta) = 
\begin{cases} 
\frac{\partial }{\partial x_i} \mu_{i,j}(x,\beta) =   2(x_j - x_i) e^{ -  \|x_i-x_j\|_2^2 + f^F_{ij}(\beta, k) + f^R_{ij}(\beta, k)} ,\\ 
\frac{\partial }{\partial x_j} \mu_{i,j}(x,\beta)  = -   2(x_j - x_i) e^{ -  \|x_i-x_j\|_2^2 + f^F_{ij}(\beta, k) + f^R_{ij}(\beta, k)},  \\
\frac{\partial }{\partial x_k} \mu_{i,j}(x,\beta) = 0, & \mbox{ for } k\neq i,j
\end{cases}
\end{equation}

\section{EKF}
\label{A}
The posterior variance is calculated keeping the Taylor local approximation $\mu(x_k,  \beta) \approx H_k x_k $  
\begin{equation}
\notag
\begin{split}
V_{k|k} &= \mathbb{E}[(x_k-\hat{x}_{k|k})(x_k-\hat{x}_{k|k})']  = \mathbb{E}[(x_k-\hat{x}_{k|k-1} -K_k(y_k - H_k\hat{x}_{k|k-1}) )(x_k-\hat{x}_{k|k-1} -K_k(y_k -H_k\hat{x}_{k|k-1}))'] \\
& = \mathbb{E}[(x_k-\hat{x}_{k|k-1} -K_k(H_k x_k + \epsilon_k -H_k\hat{x}_{k|k-1}) )(x_k-\hat{x}_{k|k-1} -K_k(H_k x_k + \epsilon_k -H_k\hat{x}_{k|k-1}))'] \\
& = \mathbb{E}[(x_k-\hat{x}_{k|k-1})(x_k-\hat{x}_{k|k-1})'] + \mathbb{E}[K_k (H_k x_k -H_k\hat{x}_{k|k-1})(H_k x_k -H_k\hat{x}_{k|k-1})'K_k'] + \mathbb{E}[K_k \epsilon_k \epsilon_k' K_k'] \\
& - \mathbb{E}[K_k H_k (x_k - \hat{x}_{k|k-1}))(x_k - \hat{x}_{k|k-1}))'] - \mathbb{E}[ (x_k - \hat{x}_{k|k-1}))(x_k - \hat{x}_{k|k-1}))'H_k' K_k'] \\
& = V_{k|k-1} + K_k H_k V_{k|k-1} H_k' K_k' + K_k R_k K_k' - K_k H_k V_{k|k-1} - V_{k|k-1} H_k'  K_k'    
\end{split}
\end{equation}
where
\begin{equation}
\notag
K_k H_k V_{k|k-1} H_k' K_k' + K_k R_k K_k' = K_k (H_k V_{k|k-1} H_k' + R_k) K_k' = V_{k|k-1} H_k' K_k' 
\end{equation}
thus
\begin{equation}
\notag
V_{k|k} = V_{k|k-1}  - K_k H_k V_{k|k-1} 
= (\mathbb{I} - K_k H_k ) V_{k|k-1}.
\end{equation}

\section{Smoother}
\label{B}
\begin{equation}
\notag
\begin{split}
\mathbb{E}\left[\mathbb{V}\left[x_{k-1} |x_{k}, y \right] | y\right] &= \mathbb{E}\left[\mathbb{V}\left[x_{k-1} |x_{k}, y_{1:k-1} \right] | y\right] \\  
&= \mathbb{E}\left[\mathbb{V}\left[x_{k-1} | y_{1:k-1} \right] - \mathbb{C}ov(x_{k-1}, x_k |y_{1:k-1}) \mathbb{V}(x_k |y_{1:k-1})^{-1} \mathbb{C}ov(x_{k-1}, x_k |y_{1:k-1})' | y\right] \\
&= \mathbb{E}\left[ V_{k-1|k-1} - B_k V_{k|k-1} B_k' | y\right] = V_{k-1|k-1} - B_k V_{k|k-1} B_k' \\
\mathbb{V}\left[\mathbb{E}\left[x_{k-1} |x_{k}, y \right] | y\right]& = \mathbb{V}\left[ \hat{x}_{k-1|k-1} + B_k (x_k - \hat{x}_{k|k-1})  | y\right] = B_k V_{k|n} B_k' \\
\end{split}
\end{equation}

\section{Maximization}
\label{C}
\subsection{Poisson component}
\begin{equation}
    \begin{split}
       Q(\beta,\Sigma)&=\sum_{tij} \mathbb{E}[-\mu_{ij}(x_k, \beta)] + \mathbb{E} [y_{ij}(k)\log(\mu_{ij}(x_k, \beta))] - \log(y_{ij}(k)!)  + C_2 =\\
        & \sum_{tij} -\mathbb{E}[e^{- d(x_i(k),x_j(k))}] e^{ f^F_{ij}(\beta, k) + f^R_{ij}(\beta, k)} + \\ 
        & + y_{ij}(k)(\mathbb{E} [- d(x_i(k),x_j(k))] + f^F_{ij}(\beta, k) + f^R_{ij}(\beta, k)) - \log(y_{ij}(k)!) + C_2 
    \end{split}
\end{equation}
Notice that adding and subtracting $y_{ij}(k)\log(\mathbb{E} [e^{- d(x_i(k),x_j(k))}])$
\begin{equation}
    \begin{split}
        &y_{ij}(k)(\mathbb{E} [- d(x_i(k),x_j(k))] + f^F_{ij}(\beta, k) + f^R_{ij}(\beta, k)) \\
        &=y_{ij}(k)(\log(\mathbb{E} [e^{- d(x_i(k),x_j(k))}]) + f^F_{ij}(\beta, k) + f^R_{ij}(\beta, k)) + \\
        & + y_{ij}(k)\mathbb{E} [- d(x_i(k),x_j(k))] - y_{ij}(k)\log(\mathbb{E} [e^{- d(x_i(k),x_j(k))}])
    \end{split}
\end{equation}
thus
\begin{equation}
\begin{split}
Q(\beta,\Sigma)&= \sum_{tij} -\mathbb{E}[e^{- d(x_i(k),x_j(k))}] e^{ f^F_{ij}(\beta, k) + f^R_{ij}(\beta, k)} + \\
&+ y_{ij}(k)(\log(\mathbb{E} [e^{- d(x_i(k),x_j(k))}]) + f^F_{ij}(\beta, k) + f^R_{ij}(\beta, k)) - \log(y_{ij}(k)!) + C_3  \\
&= \sum_{tij} - \mu_{ij}^*(x_k, \beta) + y_{ij}(k)(\log(\mu_{ij}^*(x_k, \beta)) - \log(y_{ij}(k)!) + C_3
\end{split}
\end{equation}

\subsection{Gaussian component}
\begin{equation}
\begin{split}
\label{eqn:EM}
\hat{\Sigma} &= \mathbb{E}\left[ \frac{1}{n} \sum_1^n  (x_k - x_{k-1})(x_k - x_{k-1})' \big|y_{1:n} \right] = \frac{1}{n} \sum_1^n \mathbb{E}\left[ (x_k - x_{k-1})(x_k - x_{k-1})' \big|y_{1:n} \right]  \\  
&= \frac{1}{n} \sum_1^n \mathbb{E}\left[x_k x_k'\big|y_{1:n}\right] + \mathbb{E}\left[ x_{k-1} x_{k-1}'\big|y_{1:n}\right] - \mathbb{E}\left[x_{k-1} x_{k}'\big|y_{1:n}\right] - \mathbb{E}\left[x_{k} x_{k-1}'\big|y_{1:n}\right]  \\
&= \frac{1}{n} \sum_1^n  V_{k|n} + V_{k-1|n} + B_k V_{k|n} +  V_{k|n}B_k' + (\hat{x}_{k|n} - \hat{x}_{k-1|n})(\hat{x}_{k|n} - \hat{x}_{k-1|n})'
\end{split}    
\end{equation}
\begin{equation}
\notag
\begin{split}
\mathbb{E}\left[x_k x_k'\big|y_{1:n}\right] &= \mathbb{E}\left[((x_k - \hat{x}_{k|n}) + \hat{x}_{k|n}) ((x_k - \hat{x}_{k|n}) + \hat{x}_{k|n})'\big|y_{1:n}\right] \\
&= \mathbb{E}\left[(x_k - \hat{x}_{k|n}) (x_k - \hat{x}_{k|n})'\big|y_{1:n}\right] + \hat{x}_{k|n}\hat{x}_{k|n}' = V_{k|n} + \hat{x}_{k|n}\hat{x}_{k|n}'\\
\mathbb{E}\left[x_k x_{k-1}'\big|y_{1:n}\right] &= \mathbb{E}\left[x_k \mathbb{E}\left[x_{k-1}'\big|x_k, y_{1:k-1}\right]\big|y_{1:n}\right] = \mathbb{E}\left[x_k (\hat{x}_{k|k} + B_k(x_k - \hat{x}_{k|k-1}))'\big|y_{1:n}\right] \\ 
&= \mathbb{E}\left[((x_k - \hat{x}_{k|n}) + \hat{x}_{k|n}) (\hat{x}_{k-1,k-1} + B_k((x_k - \hat{x}_{k|n}) + \hat{x}_{k|n} - \hat{x}_{k|k-1}))'\big|y_{1:n}\right] \\
&= \mathbb{E}\left[(x_k - \hat{x}_{k|n}) (x_k - \hat{x}_{k|n})'\big|y_{1:n}\right]B_k' + \hat{x}_{k|n}(\hat{x}_{k-1,k-1} + B_k( \hat{x}_{k|n} - \hat{x}_{k|k-1}))' \\
&=   V_{k|n}B_k' + \hat{x}_{k|n}\hat{x}_{k-1|n}'
\end{split}    
\end{equation}

\section{Alternative derivation of EKF}
\label{D}
The Poisson distribution can be written in the natural exponential family formulation \citep{mccullagh2018generalized}:
\begin{equation}
\notag
\begin{split}
&  p(x_k | x_{k-1}) = \frac{1}{\sqrt{2\pi}}|\Sigma|^{-1} e^{-\frac{1}{2}(x_k - x_{k-1})' \Sigma^{-1} (x_k - x_{k-1}) } \\
&  p(y_k | x_k) = c(y_k) e^{\theta' y_k - b(\theta) } \\
& b(\theta) = 1'e^\theta \\
& \theta(x_k)=  \log \mu(x_k,  \beta) \\
& \mu(x_k,  \beta) = \mathbb{E}[y_k | x_k] = \frac{\partial }{\partial \theta} b(\theta)  \\
& R_k = \mathbb{V}[y_k | x_k] =  \frac{\partial^2 }{\partial \theta^2} b(\theta).  \\
    & b(\theta) : \mathbb{R}^{p_y} \rightarrow \mathbb{R} \\
    & \theta(\mu)  = \left[\frac{\partial }{\partial \theta} b(\theta) \right]^{-1} : \mathbb{R}^{p_y} \rightarrow \mathbb{R}^{p_y}.
\end{split}    
\end{equation}
The advantage of writing the Poisson distribution in the natural exponential family form is that the further developments will be valid for any distribution of the natural exponential family. Other exponential family distributions are possible specifying differently the functions $\theta(\cdot)$ and $b(\cdot)$.
The likelihood can be then written as
\begin{equation}
\label{eqn:likComplete}
\begin{split}
L(\beta, \Sigma; y, x) = \prod_{k=1}^n
\frac{1}{\sqrt{2\pi}}|\Sigma|^{-1} e^{-\frac{1}{2}(x_k - x_{k-1})' \Sigma^{-1} (x_k - x_{k-1}) } 
c(y_k) e^{\theta' y_k - b(\theta) }
\end{split}
\end{equation}

We obtain the correction step via maximum likelihood.
The likelihood that we are treating here is  different than the one presented in (\ref{eqn:lik}).
We are taking the single likelihood contribution at time $k$ conditioned to the inference at the previous time point. Thus the marginal distribution of the latent process is substituted with its conditional distribution, i.e., the distribution that we calculated in the prediction step.
The likelihood is presented as

\begin{equation}
\label{eqn:singlellik}
l_k(x_k) = -\frac{1}{2}(x_k - \hat{x}_{k|k-1})' V_{k|k-1}^{-1}(x_k - \hat{x}_{k|k-1}) +\theta'y_k - b(\theta)
\end{equation}
were $V_{k|k-1}$ represent the variance of the latent process conditioned to $y_{k-1}$.
From a frequentist point of view (\ref{eqn:singlellik}) is a penalized likelihood, composed by the Poisson probability of the observations and a penalty term for the latent process.
In a Bayesian setting it can be considered a posterior distribution, where the penalty represents the prior distribution.
The penalty/prior regulates the smoothness of the process via the covariance matrix $\Sigma$. 
The maximization of the posterior density is equivalent to the maximization of the penalized likelihood \citep{fahrmeir1992posterior}.
We maximize this likelihood according to $x_k$, to obtain $\hat{x}_{k|k}$. This clearly is not equivalent to the conditional mean, except in case the posterior mode coincide with the posterior mean. This is true for the Gaussian density, which is not our case. The posterior is therefore approximated with the same family distribution of the prior, i.e., Gaussian, see \cite{gamerman1991dynamic} and  \cite{fahrmeir1992posterior}. Thus we are approximating the posterior mean with the posterior mode.

Using the chain rule, we take the derivative of the likelihood respect to $x_k$ and transposing it we have
\begin{equation}
\notag
\begin{split}
& \frac{\partial }{\partial x_k}l_k(x_k) = -V_{k|k-1}^{-1} (x_k - \hat{x}_{k|k-1}) + \frac{\partial \mu(x_k,  \beta)}{\partial x_k}'\frac{\partial \theta(\mu)}{\partial \mu}(y_k - \frac{\partial }{\partial \theta} b(\theta)). 
\end{split}
\end{equation}
A first order Taylor expansion is applied on the mean of $y_k$
\begin{equation}
\label{eqn:taylor}
\frac{\partial }{\partial \theta} b(\theta) = \mu(x_k,  \beta) = \mu(\hat{x}_{k|k-1}) + \frac{\partial \mu(x_k,  \beta) }{\partial x_k} (x_k-\hat{x}_{k|k-1})
\end{equation}
obtaining
\begin{equation}
\notag
\frac{\partial }{\partial x_k}l_k(x_k) = -V_{k|k-1}^{-1}(x_k - \hat{x}_{k|k-1}) + \frac{\partial \mu(x_k,  \beta)}{\partial x_k}'\frac{\partial \theta(\mu)}{\partial \mu}\left(y_k - \mu(\hat{x}_{k|k-1}) - \frac{\partial \mu(x_k,  \beta) }{\partial x_k} (x_k-\hat{x}_{k|k-1})\right) 
\end{equation}
Setting $\frac{\partial }{\partial x_k}l_k(x_k)=0$ and rearranging the members of the equation we have
\begin{equation}
\notag
x_k = \hat{x}_{k|k-1} + \left[ V_{k|k-1}^{-1} + \frac{\partial \mu(x_k,  \beta) }{\partial x_k}'\frac{\partial \theta(\mu)}{\partial \mu}\frac{\partial \mu(x_k,  \beta)}{\partial x_k}  \right]^{-1} \left[ \frac{\partial \mu(x_k,  \beta) }{\partial x_k}'\frac{\partial \theta(\mu)}{\partial \mu} \right] \left(y_k - \mu(\hat{x}_{k|k-1}) \right).
\end{equation}
We evaluate the derivatives at $\hat{x}_{k|k-1}$ and use the property that the second derivative of $b(\theta)$ is equal to the variance of $y_k|x_k$. Since $x_k$ is unknown, we approximate it with $\hat{x}_{k|k-1}$.
\begin{equation}
\label{eqn:R}
\frac{\partial \theta(\mu) }{\partial \mu} \big|_{\hat{x}_{k|k-1}} = \left(\frac{\partial^2 b(\theta) }{\partial \theta^2} \right)^{-1} \big|_{\hat{x}_{k|k-1}}  = \mathbb{V}(y_k |x_k) ^{-1} \big|_{\hat{x}_{k|k-1}} = R_k^{-1} .
\end{equation}
Setting
\begin{equation}
\notag
 \frac{\partial \mu(x_k,  \beta) }{\partial x_k} \big|_{\hat{x}_{k|k-1}} = H_k\\
\end{equation}
and considering that
\begin{equation}
\notag
 \mu(\hat{x}_{k|k-1})  = H_k \hat{x}_{k|k-1}
\end{equation}
we obtain the update

\begin{equation}
\notag
\begin{split}
 \hat{x}_{k|k} &= \hat{x}_{k|k-1} + [ V_{k|k-1}^{-1} + H_k' R_k^{-1}H_k  ]^{-1} [  H_k' R_k^{-1} ] (y_k - H_k \hat{x}_{k|k-1} ) \\
& = \hat{x}_{k|k-1} + K_k (y_k - H_k \hat{x}_{k|k-1} ).
\end{split}
\end{equation}
The last equation comes under the name of Information Filter. $V_{k|k-1}^{-1}$ is the information matrix on $x_k$ given $y_{1:k-1}$, $H_k'R_k^{-1}H_k$ is the information on $x_k$ contributed by the last observation $y_{k}$ and the sum of the two is the information on $x_k$ given $y_{1:k}$. Considering that the numerator $[  H_k' R_k^{-1} ] (y_k - H_k \hat{x}_{k|k-1} )$ is the first derivative, the correction step has the form of a single Fisher scoring step \citep{fahrmeir1992posterior}.
The formula of the filter can be rearranged in the following way
\begin{equation}
\notag
\begin{split}
 K_k &= ( V_{k|k-1}^{-1} + H_k' R^{-1}H_k  )^{-1} H_k' R_k^{-1} = ( V_{k|k-1}^{-1} + H_k' R^{-1}H_k  )^{-1}  H_k' R_k^{-1} (R_k + H_k V_{k|k-1} H_k')(R_k + H_k V_{k|k-1} H_k')^{-1} \\
& =( V_{k|k-1}^{-1} + H_k' R_k^{-1}H_k  )^{-1} (V_{k|k-1}^{-1} + 'R_k^{-1} ) V_{k|k-1} H_k' (R_k + H_k V_{k|k-1} H_k')^{-1} \\
& = V_{k|k-1} H_k' (R_k + H_k V_{k|k-1} H_k')^{-1}
\end{split}
\end{equation}
obtaining the filtering matrix for the EKF.

\end{document}